# Quantum Coherent States and Path Integral Method to Stochastically Determine the Anisotropic Volume Expansion in Lithiated Silicon Nanowires

**Donald C. Boone**

Nanoscience Research Institute, 4423 Lehigh Road Suite 371, College Park, MD 20740, USA; db2585@caa.columbia.edu

**Abstract:** This computational research study will analyze the multi-physics of lithium ion insertion into a silicon nanowire in an attempt to explain the electrochemical kinetics at the nanoscale and quantum level. The electron coherent states and a quantum field version of photon density waves will be the joining theories that will explain the electron-photon interaction within the lithium-silicon lattice structure. These two quantum particles will be responsible for the photon absorption rate of silicon atoms that are hypothesized to be the leading cause of breaking diatomic silicon covalent bonds that ultimately leads to volume expansion. It will be demonstrated through the combination of Maxwell stress tensor, optical amplification and path integrals that a stochastic analyze using a variety of Poisson distributions that the anisotropic expansion rates in the <110>, <111> and <112> orthogonal directions confirms the findings ascertained in previous works made by other research groups. The computational findings presented in this work are similar to those which were discovered experimentally using transmission electron microscopy (TEM) and simulation models that used density functional theory (DFT) and molecular dynamics (MD). The refractive index and electric susceptibility parameters of lithiated silicon are interwoven in the first principle theoretical equations and appears frequently throughout this research presentation, which should serve to demonstrate the importance of these parameters in the understanding of this component in lithium ion batteries.

## 1. Introduction

The research work that will be presented is a continuation from a study that examined how energy was transformed within an electron flux that transverse a cubic silicon crystal lattice in the opposing direction of lithium ion diffusion [1]. The previous study was a computational model to examine how a surplus of electron charge could lead to an exponential increase in applied electromagnetic energy that could possibly result in enough energy to sever the covalent bonds between silicon atoms that was hypothesis to be the genesis of the anisotropic volume expansion witness during lithium ion insertion. This work used an interdisciplinary approach within physics with quantum mechanics serving as the main mathematical framework. The preceding study, which is the edifice of this current work, uses first principle theories that are accepted throughout physics. This methodology will continue in this paper. It is the goal at the end of this research study that we will come closer to the elusive explanation of the electrochemical kinetics of lithiated silicon nanowires.

The research body of knowledge for lithiated silicon anode materials has been exclusively focused on lithium ion diffusion process. The research work by Liu et al. anisotropic volume expansion of lithiated silicon nanowires was studied employing transmission electron microscope (TEM) and electron diffraction pattern (EDP) [2]. In this study, a morphology evolution of the lithiated silicon nanowire started from a pristine crystalline silicon (c-Si) nanowire at 155 nm in diameter prior to lithium insertion to a 17% diameter increase in the <111> direction and a 170% increase in diameter of 485 nm was measured in the <110> direction after full lithiation. In addition, a crack along the longitudinal direction of <112> was detected. A similar results were performed by Yang et al. by utilizing a chemomechanical finite element model to simulate several crystallography orientation-dependent anisotropic volume expansions of over 300%, with increases initiating at the interfacial reaction fronts of lithiated silicon nanowire models [3]. The research study performed by Cubuk et al. used the kinetic Monte Carlo (kMC) method to simulate the lithium atoms insertion into silicon nanowire. The expansion rates were calculated seven times faster in the <110> direction than the <111> direction [4]. This gave the expanded lithiated silicon nanowire a described "dumbbell cross section" shape that resembled the Cassini oval curve geometry. The work drawn from Jung et al. molecular dynamics/density functional theory (MD/DFT) simulation was created to study the atomistic behavior of the two-phase interfacial reaction front barrier that separates the c-Si and LixSi material [5]. All of these works support the findings that the volume of silicon nanowires during lithium atom/ion insertion will increase non-isotopically by nature.

For the purpose of continuity, our theoretical apparatus will be briefly re-examined as it was presented in our previous work. Prior to the beginning of the lithiation process, the individual lithium atoms are ionized reducing them to the constitutive particles of lithium ions and free electrons [6]. A constant voltage of 2 V is applied to an electric series circuit in order for the lithiation process to begin. The electrons and lithium ions will enter the silicon nanowire at opposing ends and therefore travel in opposite directions (Figure 1). When the lithiation begins, this initiates a process of transforming the silicon from c-Si to an amorphous lithiated silicon (a-LiSi) matrix [7,8]. The free electrons or electron flux varies and increases with the continue diffusion of lithium ions within the silicon nanowire [9]. Since the electrons are moving charge particles, they are the source of the applied quantized electromagnetic field. The geometric model that will be the basis of our mathematical framework is a diamond crystal silicon lattice, which is composed of eight silicon atoms (Figure 2). The silicon lattice is fully lithiated with 30 lithium ions [10,11].

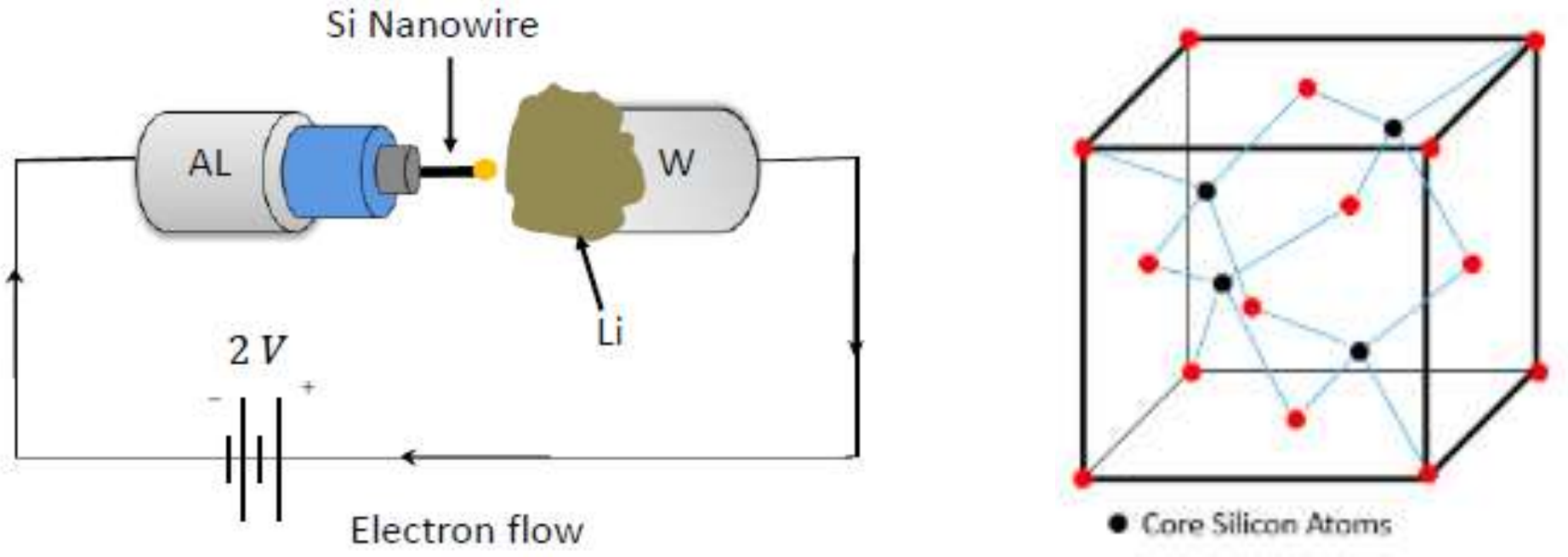

(**a**) (**b**)

**Figure 1.** (**a**) In situ experimental arrangement for a solid electrochemical cell using lithium metal counter electrode (W); (**b**) Silicon is a diamond crystalline cubic structure made up of tetrahedral molecules with its hybridized $sp^3$ orbitals within their valence shells filled with covalent bonding electrons from neighboring silicon atoms. Reprinted from AIP Advances 6, 125027 (2016) American Institute of Physics. Copyright by the author under the terms and conditions of the Creative Commons Attribution (CC BY 4.0).

There will be six theories that will be presented based on first principle, each describing a part of the mathematical framework that leads to the anisotropic volume expansion in this model. Each theory operates simultaneously and interdependent of each other, in fact some of the equations and parameters, especially refractive indices and electric susceptibilities, will be found repeatedly in the theories. The first theory defines the state of the net electric charge in terms of a series of possible Poisson distribution curves. The second theory that will be presented is a recreation of the work previously done by the author, Maxwell stress with optical amplification. The third theory will introduce the type of quantum harmonic oscillator that will define the energy of each individual electron within the model. The fourth theory will introduce coherent states that define the probability of the individual electrons in the quantum harmonic oscillator. The photon density wave theory will incorporate a quantum electromagnetic field to examine the photon absorption rate of the silicon atoms, which will the fifth theory presented. The sixth and final theory will be the path integral method as originally theorized by Richard Feynman which will ultimately calculate the expansion rates in each of the orthogonal directions and predict the increase change in volume along with describing the varies geometric cross-sectional areas of the silicon nanowire.

## 2. Poisson Distribution of the Negative Charge Differential ($n_c$)

As mentioned previously, negative free electrons and positive lithium ions enter into our computational model from opposing directions. The electric charge difference between the two constitutive mass particles, where the electrons are always greater or equal in number to the lithium ions in our model, will be known as the negative charge differential $n_c$ which exist within the conduction band of the quantum harmonic oscillator. The state of the negative charge distribution within the conduction band is defined by a Poisson distribution called the negative charge differential probability $N_{pq}$.

$$N_{pq} = \sum_{\bar{n}_c=1}^{\infty} \sum_{n_c=1}^{\infty} \frac{\bar{n}_c{}^{n_c} exp(-\bar{n}_c)}{n_c!} \tag{1}$$

where $p = \bar{n}_c$ and $q = n_c$. The variable $\bar{n}_c$ is the average negative charge differential which is simply the mean of $n_c$ during a given time interval. The $\bar{n}_c$ parameter will be the main independent variable in this research work. The negative charge differential $n_c$ will be thought of as varying at each moment in time. Any particular average $\bar{n}_c$ will be defined by the Poisson curve from $0 \leq n_c \leq \infty$ with the peak value of any curve being the mean value $\bar{n}_c$. Each $n_c$ is assigned a probability between 0 to 1. Figure 2 displays a sample of the family of Poisson curves.

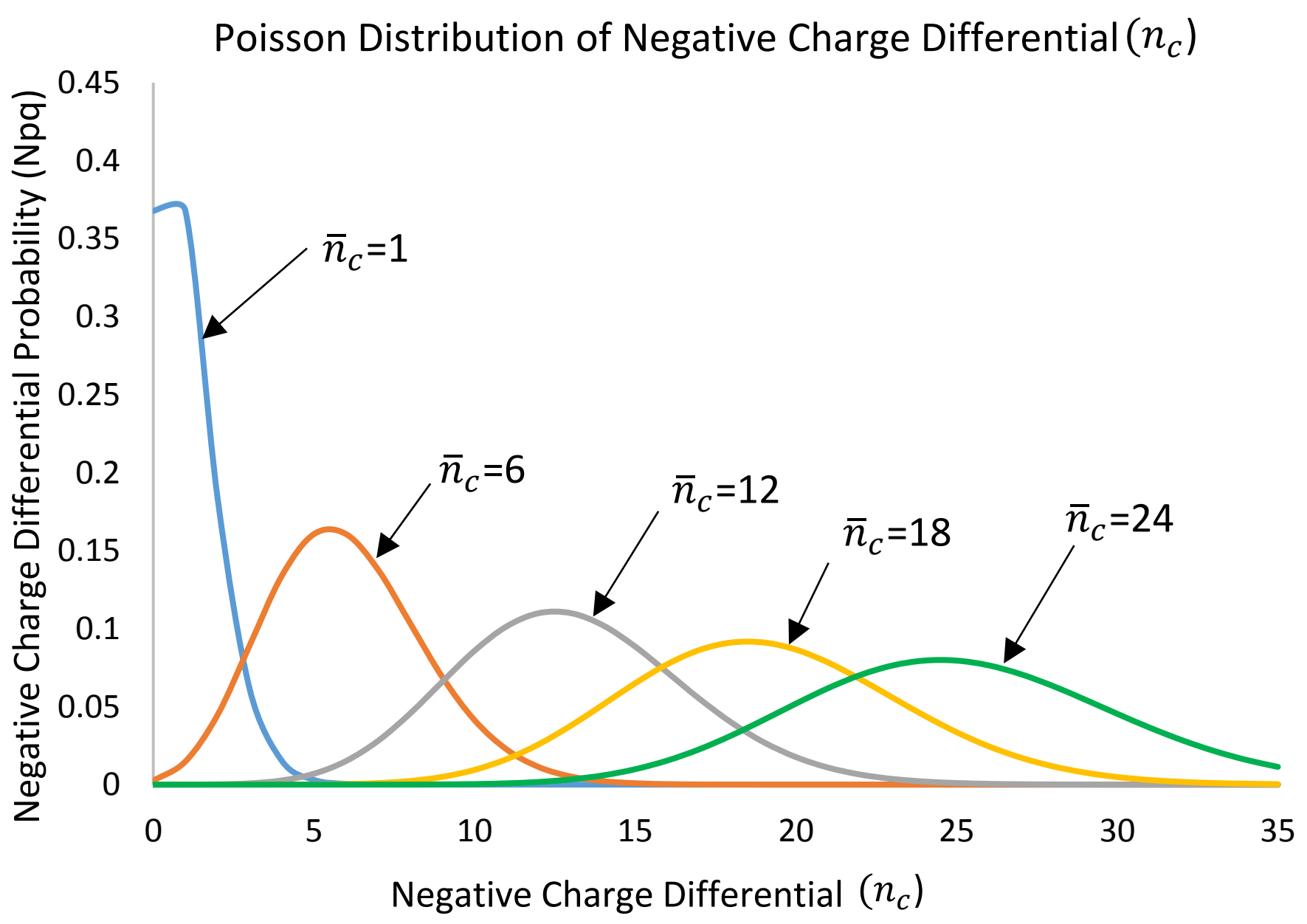

**Figure 2.** The net difference between negative electrons and positive lithium ions is called the negative charge differential $n_c$ and it is defined by a set of Poisson distribution curves with each curve having a peak value being the average of the sum of $n_c$ for that particular curve define as $\bar{n}_c$.

The $n_c$ variable can be thought of as being analogue to the electric charge variable found in Butler-Volmer equation, which describes electrochemical kinetics by defining the macroscale current density within a battery.

## 3. Maxwell Stress

For this research work, there will be a special notion that will be used throughout this study to indicate the orthogonal directions that are essential element in the presentation of this paper. As an example, the mathematical variables and functions that have directional characteristics will have subscripted notations that will indicate which orthogonal direction is being represented:

$$A_{ij} = A_{\langle Orthogonal\ direction\rangle} \quad \text{i=j=1 or 2 or 3;}$$
$$A_{11} = A_{\langle 110\rangle}$$
$$A_{22} = A_{\langle 111\rangle}$$
$$A_{33} = A_{\langle 112\rangle}$$

The negative charge differential $n_c$ resides within the conduction bands of the lithiated silicon lattice. In order to construct the minimum conduction band, a Bloch function based on electron scattering theory is used

$$\mathrm{u_c(r)} = \mathrm{e^{ikr}} + \frac{1}{\mathrm{kr}}\mathrm{e^{i(\delta+kr)}}\sin\delta + \frac{3\mathrm{z}}{\mathrm{kr^2}}\mathrm{e^{i(\delta+kr)}}\sin\delta \tag{2}$$

with position vectors **r** and **z** as variables [12]. The wavenumber k is the expectation value of <k> based on the silicon atoms' ground state wave function within the c-Si lattice prior to the beginning of lithiation. The silicon ground state wave function was calculated using Slater determinant of the four core silicon atoms in c-Si lattice as shown in Figure 2. The phase shift $\delta(\mathrm{k,r})$ is define as function of both <k> and position vector **r**. The Bloch function consist of three terms, the first term is known as the incident wave function and the second and third term collectively is called the scattering wave function. When an electron is traveling in the conduction band, the first term of the wave function describes the electron from a point before impact with the oncoming atom (incident wave function) and the second and third term of the wave function describes the electron after the impact with the colliding atom (scattering wave function). The applied electric field $\vec{E}_{ij}$ and magnetic field $\vec{B}_{ij}$ are defined as

$$\vec{E}_{ij} = \mathrm{iC}_{\mathrm{E}ij}\frac{\bar{\mathrm{h}}^2(3\pi^2\bar{n}_\mathrm{c})^{\frac{2}{3}}\mathrm{v}_{DOS}}{4\mathrm{n_v e}m_{eff}}[\mathrm{u_c}\nabla\mathrm{u_c^*} - \mathrm{u_c^*}\nabla\mathrm{u_c}]\exp^{\frac{\gamma_{ij}r}{2}} \tag{3}$$

$$\vec{B}_{ij} = \mathrm{iC}_{\mathrm{B}ij}\frac{\bar{\mathrm{h}}(3\pi^2\bar{n}_\mathrm{c})^{\frac{1}{3}}\mathrm{v}_{DOS}}{2\mathrm{n_v e}}[\mathrm{u_c}\nabla\mathrm{u_c^*} - \mathrm{u_c^*}\nabla\mathrm{u_c}]\exp^{\frac{\gamma_{ij}r}{2}} \tag{4}$$

$$m_{eff} = m_{ij} = \frac{r_g \mathrm{e}\vec{B}_{ij}}{v_d} \tag{4a}$$

where e is the electric charge of an electron, ħ is the Planck's constant, $\mathrm{n_v}$ is the electron density of the maximum valence band, $\mathrm{v}_{DOS}$ is defined as the density of state volume and $\gamma_{ij}$ is the optical amplification, that will be explained later in this study, $\mathrm{m_{eff}}$ is the electron effective mass due to the cyclotron resonance of the magnetic field induced by the silicon atoms, the radius of gyration $r_g$ is the helix radial motion of the electron as it travels under the influence of $\vec{B}_{ij}$, the drift velocity $v_d = v_{ij}$ of the electrons due to the 2-V source applied to the electric circuit. There is an electron cyclotron effective mass $m_{eff} = m_{ij}$ assigned to each of the three orthogonal directions of <110>, <111> and <112>. The fact that $m_{eff}$ within the silicon matrix is dependent on the crystallography directions has been well researched and studied for many years [13]. A hypothesis of this computational research is, that because of the anisotropic expansion of the lithiated silicon nanowire, the cyclotron effective mass $m_{eff}$ has different values in each of the orthogonal directions. The coefficients $\mathrm{C}_{\mathrm{E}ij}$ and $\mathrm{C}_{\mathrm{B}ij}$ are for the electric and magnetic fields that will allow these fields to be a solution to Maxwell equations. The energy density $\mathbb{E}_{ij}$ of the applied electromagnetic (EM) field is defined as

$$\mathbb{E}_{ij} = \frac{1}{2}\left[\varepsilon_{rij}\vec{E}_{ij}^{\ 2} + \frac{1}{\mu_{ij}}\vec{B}_{ij}^{\ 2}\right] \quad (5)$$

with $i$ and $j$ are indices with values 1,2 and 3, $\varepsilon_{rij} = \varepsilon_o\varepsilon_{ij}$ is the relative electric permittivity which is the constitutive property that defines how the dielectric material affects an applied electric field and $\varepsilon_o$ is the vacuum electric permittivity. The relative magnetic permeability $\mu_{ij}$ is the constitutive property that defines the amount of magnetization a material has in response to an applied magnetic field. The magnetization is the magnetic moments per unit volume which is in essence the magnetic field that is regenerated by the spin of the individual electrons in a unit volume. In this application, the magnetic field is negligible which results in $\mu_{ij}$ being defined as unity or one. Equation (5) is also known as the Maxwell stress tensor [14,15]. When $i = j$, the three orthogonal directions that were discussed in the introduction, namely <110>, <111> and <112>, will be given by the matrix elements $\varepsilon_{11}$, $\varepsilon_{22}$ and $\varepsilon_{33}$ , respectively and are called the electric susceptibilities $\varepsilon_{ij}$ for lithiated silicon. For each of these directions we will designate a matrix element on the principle diagonal of the dielectric tensor [16,17].

$$\varepsilon_{ij} = \begin{bmatrix} \varepsilon_{11} & \varepsilon_{12} & 0 \\ \varepsilon_{21} & \varepsilon_{22} & 0 \\ 0 & 0 & \varepsilon_{33} \end{bmatrix}^{-1} \quad (6)$$

$$\varepsilon_{11} = \frac{\omega_p^2 X_{11}}{\omega_{\gamma 11}^2 - \omega_{o11}^2} \quad \varepsilon_{22} = \frac{\omega_p^2 X_{22}}{\omega_{\gamma 22}^2 - \omega_{o22}^2}$$
$$\varepsilon_{33} = \frac{\omega_p^2 X_{33}}{\omega_{\gamma 33}^2} \quad \varepsilon_{12} = -\varepsilon_{21} = \frac{\omega_{o12}}{\omega_{\gamma 12}} \frac{\omega_p^2 X_{12}}{\omega_{\gamma 12}^2 - \omega_{o12}^2} \quad (7)$$

$$\omega_{\gamma ij} = \frac{a^3}{2\hbar}\left[\varepsilon_{rij}\vec{E}_{ij}^{\ 2} + \frac{1}{\mu_{ij}}\vec{B}_{ij}^{\ 2}\right] \quad (8)$$

$$\omega_{oij} = \left[\frac{4\pi\varepsilon_0 \mathrm{M}_{ij} a_{Si}^3 \vec{E}_{ij}}{Z_{Si} e m_{eff}}\right]^{\frac{1}{2}} \quad (9)$$

$$\omega_p = \frac{1}{\hbar}\langle \Psi_{Li}^{g*} | \mathrm{H_p} | \Psi_{Li}^{g} \rangle \quad (10)$$

$$\chi_{ij} = \frac{e^2 \rho \omega_f}{\hbar a^3} \left| \langle \Psi_{Li}^{e} | r_{ij} | \Psi_{Li}^{g} \rangle \right|^2 \left[\frac{1}{(\omega_p - \omega_{\gamma ij})} + \frac{1}{(\omega_p + \omega_{\gamma ij})}\right] \quad (11)$$

$$\Psi_{Li}^{e} = \frac{\langle \Psi_{Li}^{m} | H_p | \Psi_{Li}^{g} \rangle}{(E_o - E_m)} \Psi_{Li}^{m} \quad (12)$$

$$\mathrm{H_p} = -\frac{\mathrm{e}}{2\, m_{eff}} \mathbf{B} \cdot \hat{\mathbf{L}} + \frac{\mathrm{e}^2}{8 m_{eff}} [\mathbf{B}^2 \mathbf{r}^2 - (\mathbf{B} \cdot \mathbf{r})^2] + \frac{\mathrm{e}^2 \mathbf{E}^2}{2\, m_{eff} \omega_{ij}^{\ 2}} \quad (13)$$

where $\omega_p$ is the excited state energy per $\hbar$ experienced by the lithium ion when they are in excitation, $\omega_{\gamma ij}$ is the energy per $\hbar$ of the applied electromagnetic field, $a^3$ is the silicon lattice cube volume of Figure 1 and, $\omega_{oij}$ is the resonant angular frequency of the silicon atoms. Analyzing the parameters of Equation (9), $\mathrm{M}_{ij}$ is the elastic modulus tensor for silicon, $a_{Si}$ is the silicon atomic radius and $Z_{Si}$ the atomic number of silicon [18]. The electric susceptibilities for lithium $\chi_{ij}$ is defined by the ground state lithium ion $\Psi_{Li}^{g}$, the excited state lithium ion $\Psi_{Li}^{e}$ wave functions, the resistivity $\rho$ of lithium ions within the silicon cubic lattice and the collision frequency $\omega_f$ of electrons are defined as the number of collisions per unit time an electron has between collisions with lithium-silicon particles. The excited state wave function for lithium $\Psi_{Li}^{e}$ is constructed by using time-independent perturbation theory defined in Equation (12). The orthogonal wave function is $\Psi_{Li}^{m}$ and $E_o$ and $E_m$ are the ground state energies for $\Psi_{Li}^{g}$ and $\Psi_{Li}^{m}$ respectively, $\omega_{ij}$ is the angular momentum of the electron as it travels in the conduction band and will be further defined in the review of quantum harmonic oscillators later in this study. The perturbed Hamiltonian $H_p$ of Equation (13), where $\hat{\mathbf{L}}$ is the angular momentum operator, simulates the electromagnetic energy added to the silicon nanowire the moment lithiation process begins.

## 4. Optical Amplification

The applied electromagnetic field that is composed of photons, splits the energy levels within the lithium ions due to the applied electric field (Stark Effect) and by the applied magnetic field (Zeeman Effect). When photons are absorbed by a lithium ion, the ion experiences an excitation that transitions the lithium ion from the ground state to excited state. The electrons in the excited lithium ion transition to a higher discrete energy level. Once the lithium ion transitions to an elevated energy level, it is subjected to the spontaneous emission process as displayed in Figure 3.

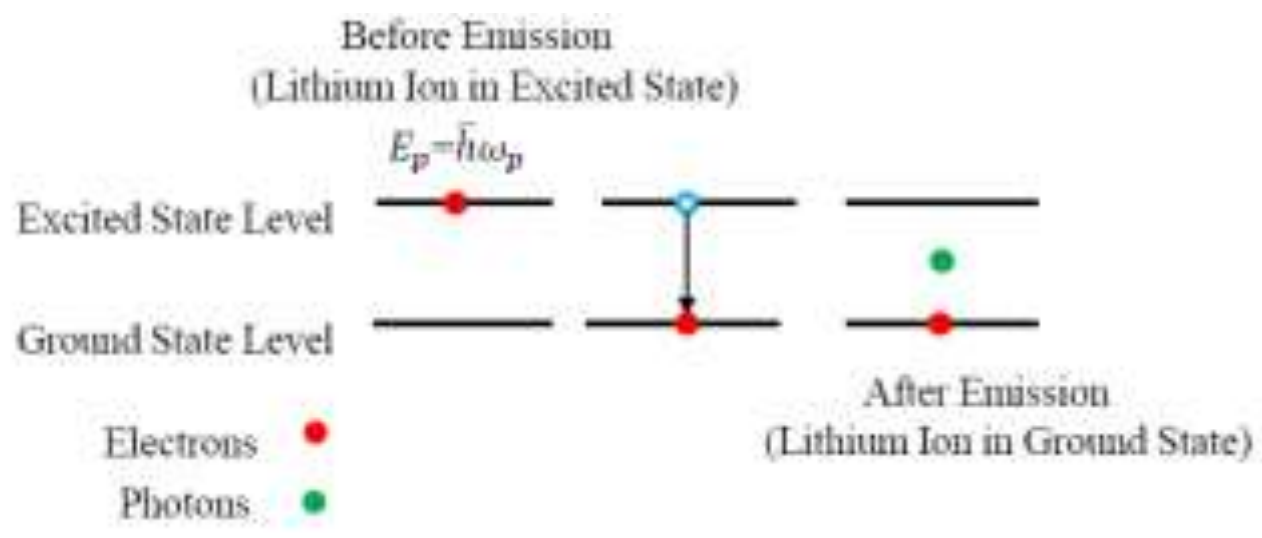

**Figure 3.** Spontaneous Emission. Reprinted from AIP Advances 6, 125027 (2016) American Institute of Physics. Copyright by the author under the terms and conditions of the Creative Commons Attribution (CC BY 4.0).

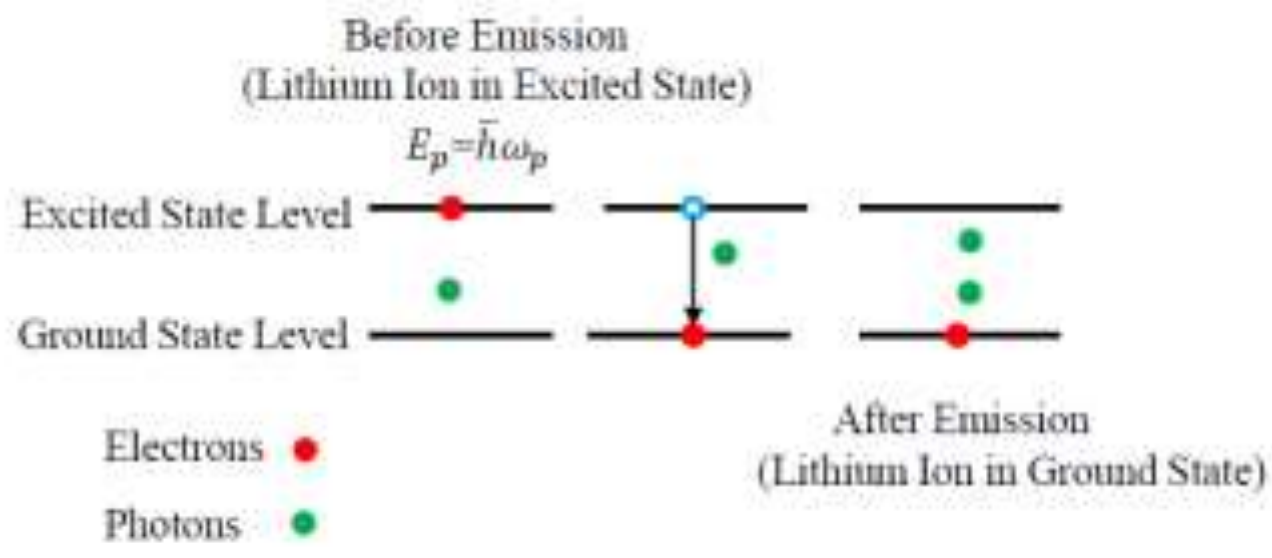

**Figure 4.** Stimulated Emission. Reprinted from AIP Advances 6, 125027 (2016) American Institute of Physics. Copyright by the author under the terms and conditions of the Creative Commons Attribution (CC BY 4.0).

Lithium ions during the diffusion process increase the density ratio of lithium to silicon atoms define by $x$ in $Li_xSi$ [19]. The increase in $x$ signifies that the lithium ions are increasing in number within the silicon lattice [20]. With the continuation of photons absorbing into the diffused lithium ions causing excitation and at the same time the diffusion process causes an increase in lithiated silicon density, the total atomic system in our lithiated silicon lattice model experiences population inversion which is define as a majority of atoms or ions being in the excited state. When the lithium ions are in such a state with photons being transmitted and absorbed within this dense lithium-silicon particle matrix, populated inversion is the prelude to the stimulated emission process (Figure 4). Stimulated emission occurs when an incoming photon interacts with a lithium ion in the excited state inducing it to transition an electron to the ground state emitting a photon that is approximately of the same frequency, phase and direction of the incoming photon. These photons, which are in the electromagnetic mode and are analogous to oscillating waves, are said to be in a state of coherence. The photons interfere with each other constructively instead of destructively. The start of stimulated emission, with the majority of lithium ions in population inversion, causes the electromagnetic intensity ($I$) to increase exponentially. The initial electromagnetic intensity $I_o$ is defined as

$$I_o = \frac{\vec{E}_{ij} \times \vec{B}_{ij}}{\mu_o \bar{h} \omega_{ij} c} \quad (14)$$

where $c$ is the speed of light and $\mu_o$ is the magnetic permeability within a vacuum. This exponential increase in $I_o$ is defined by a group of equations that includes Einstein coefficients that are used in spontaneous and stimulated absorption and emission rates:

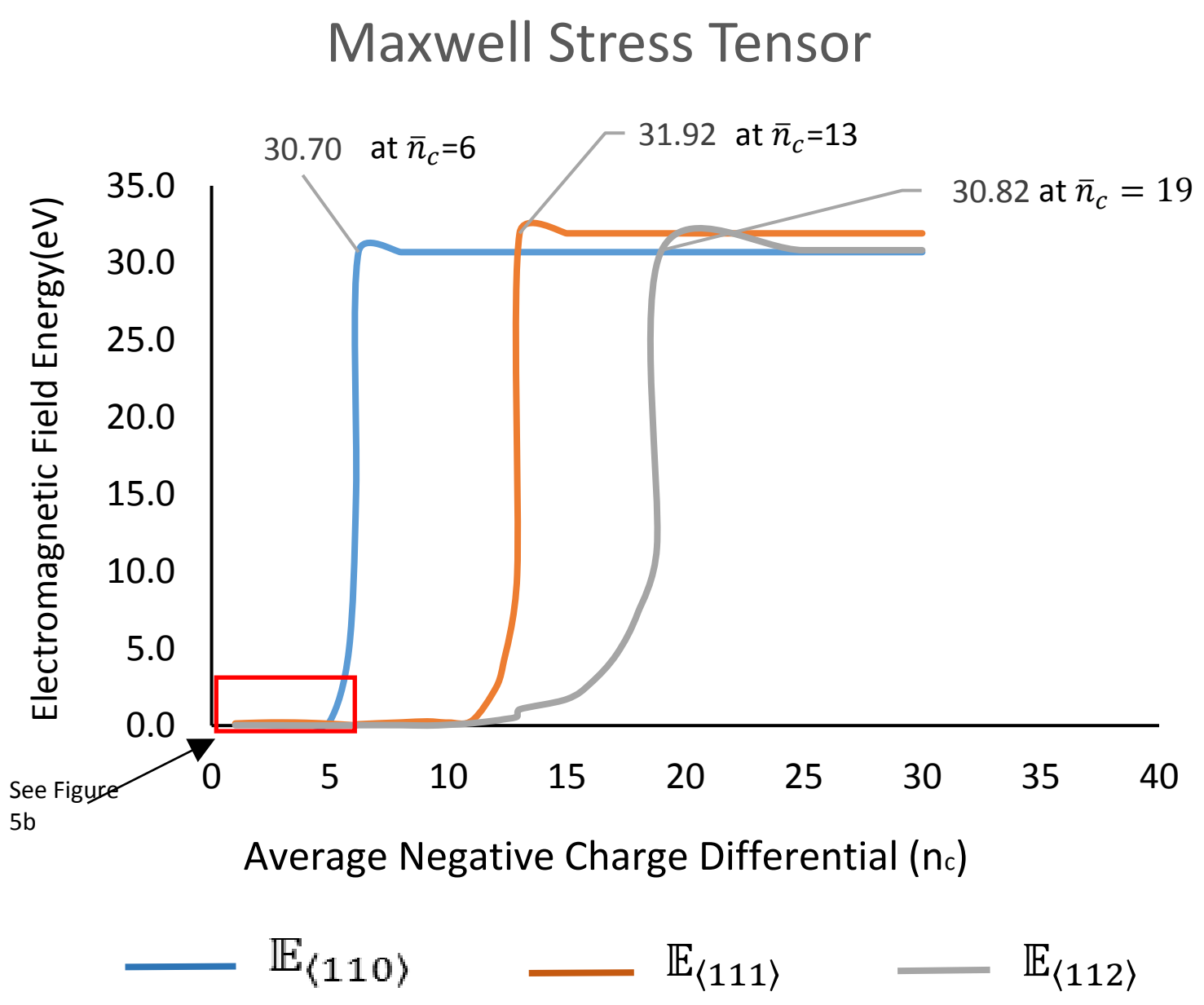

(**a**)

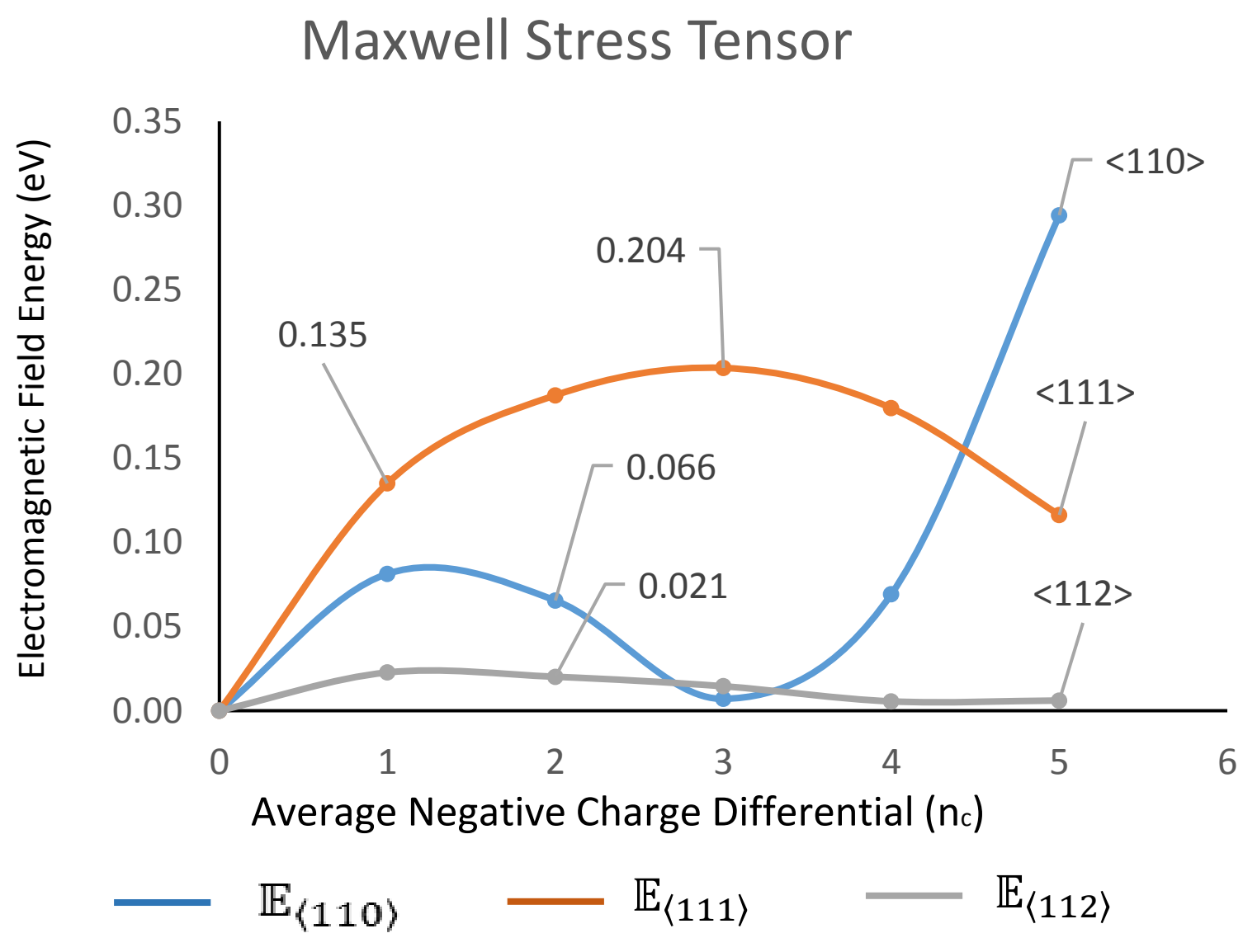

(**b**)

**Figure 5.** (**a**) The above charts demonstrate the photonic energy of the applied electromagnetic field for the <110>, <111> and <112> orthogonal directions versus the average negative charge differential $\bar{n}_c$; (**b**) The oscillatory nature of the electromagnetic field energies is derived from the wavelike nature of the electric and magnetic fields [21]. Reprinted from AIP Advances 6, 125027 (2016) American Institute of Physics. Copyright by the author under the terms and conditions of the Creative Commons Attribution (CC BY 4.0)

$$\mathrm{I}(\mathrm{r},\mathrm{t}) = \mathrm{I}_{\mathrm{o}}\exp^{\gamma_{ij}\mathrm{r}} \tag{15}$$

$$\gamma_{ij}(\mathrm{r},\mathrm{t}) = \sigma_{21ij} \cdot \Delta N_{21} \tag{16}$$

$$\Delta N_{21} = (N_2 - N_1) \tag{17}$$

$$\sigma_{21ij} = A_{21ij}\frac{\lambda^2}{8\pi n_{ij}^2}g(\omega) \tag{18}$$

$$A_{21ij} = \frac{4e^2\omega_p^3}{3\varepsilon_o\hbar c^3}\left|\langle\Psi_{Li}^g|r_{ij}|\Psi_{Li}^e\rangle\right|^2 \tag{19}$$

where $\mu_{ij} = 1$. Collectively these equations describe the process called *optical amplification* [22].

In Equation (17), $\Delta N_{21}$ is the difference of the number of excited state atoms/ions $N_2$ and the number of ground state atoms/ions $N_1$ within our model. For this study we select only the lithium ions to be in the excited state, $N_2$ = 30 and only the silicon atoms to be in the ground state $N_1$ =8. Therefore, a ratio of $x$ = $N_2/N_1$ equals 3.75, which is the same value of $x$ in $Li_xSi$ and at which the silicon diamond cubic lattice in our model is considered to be fully lithiated [23]. Equation (18) is the stimulated emission cross section area $\sigma_{21ij}$ which is defined by the Einstein $A$ Coefficient $A_{21ij}$ [24], the spectral line shape function $g(\omega)$, wavelength of the photon emitted is defined by $\lambda$ and the refractive index $n_{ij}$ of the electromagnetic field which is defined as

$$n_{ij} = \sqrt{\varepsilon_{ij}\mu_{ij}} \tag{20}$$

The electromagnetic field increases with the magnitude of $\exp^{\frac{\gamma_{ij}r}{2}}$ as noted in Equations (3) and (4). As previously mentioned, the variable $n_c$ is defined as the negative charge differential within the quantum harmonic oscillator conduction bands per unit volume which is the difference between the electrons that are traveling in the conduction bands and the number of positively charged lithium ions within silicon cubic lattice model. Theoretically, there is no applied electromagnetic field, if there is an equal number (or if there are equal numbers of) of electrons and lithium ions within the model. When the number of electrons is greater than the lithium ions an applied EM field is created. When the electrons enter the silicon nanowire, at first the electrons travel through c-Si before making contact with lithium ions. At this point the electric and magnetic fields are extremely weak. The electric field is of the order of $10^{-15}$ and the magnetic field is $10^{-22}$. The energy that is stored in the applied electromagnetic field is of the magnitude of $10^{-44}$ eV. However, when electrons cross the interfacial reaction front that separates c-Si and $Li_xSi$ the EM field increases by the order of $10^{25}$ due to the metallic properties of lithium. When electrons and lithium ions make contact and $n_c$ is non zero, the EM field diverges into the three orthogonal components of <110>, <111> and <112> as mentioned earlier in this study (Figure 5a,b). The three electric fields are of the order of $10^9$. However, the magnetic field is of the order of $10^0$ and thus its contributions to the EM field are negligible. When $\bar{n}_c = 6$ in the <110> direction, there is a great surge in the applied EM energy of the magnitude of 30.7 eV due to the Maxwell stress tensor. When the applied EM field stores this much energy, the individual photons that comprise the quantized electromagnetic field are energetic enough to break covalent bonds between silicon atoms [23]. Similarly, in the other orthogonal directions of <111> and <112>, there are surges of approximately 30 eV of EM energy, however at $\bar{n}_c$ = 13 and 19 respectively. The surges in energy in the EM field are caused by optical amplification, which is due to stimulated emission. In addition, as the electromagnetic field increases the refractive indices, $n_{ij}$ in all three orthogonal directions decrease, which contributes to the overall amplification of the EM field and EM energy (Figure 6a,b).

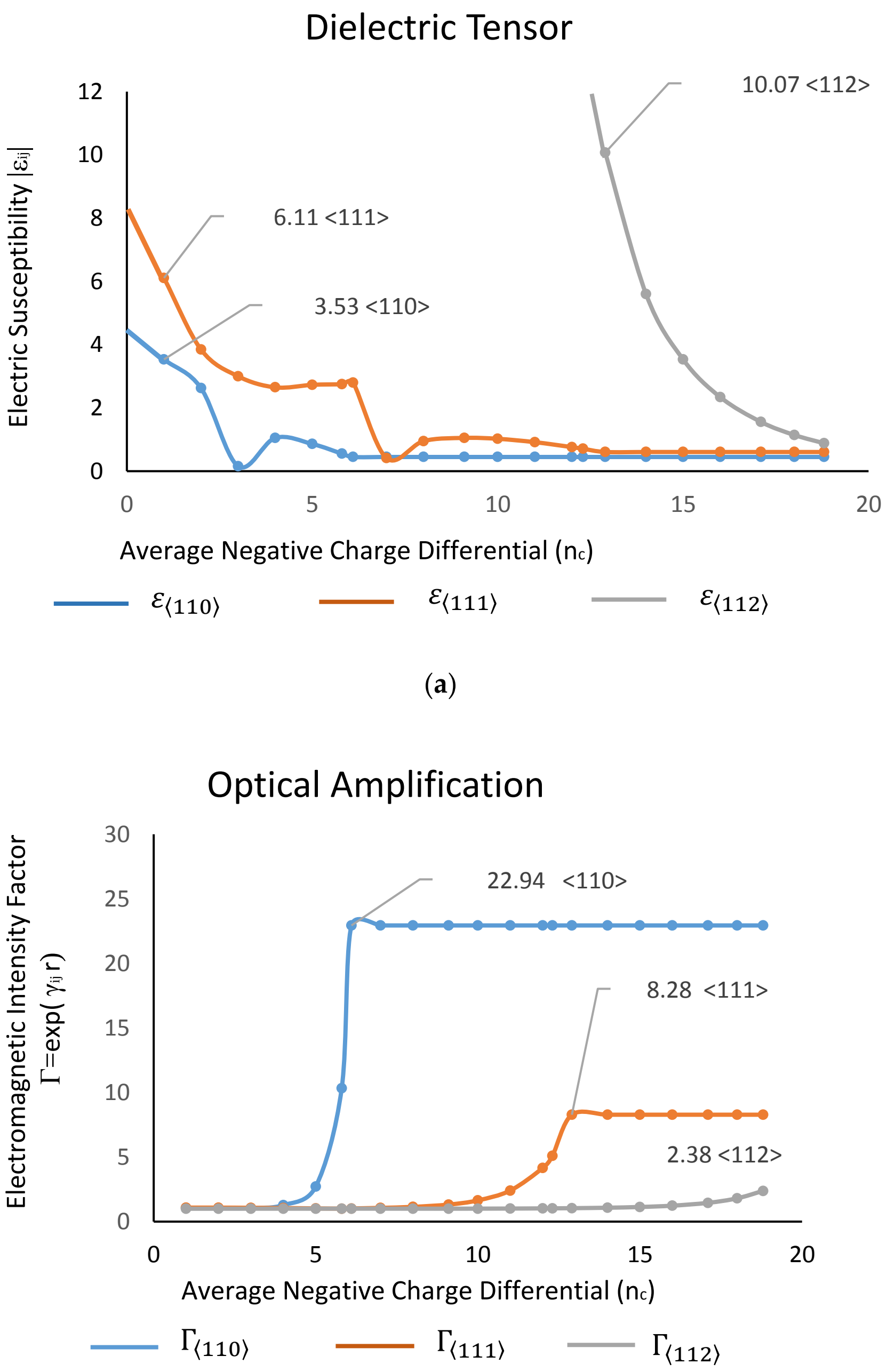

(**b**)

**Figure 6.** (**a**) The electric susceptibility $\varepsilon_{ij}$ trends downwards in all three directions as $\bar{n}_c$ increases; (**b**) The electromagnetic intensity is much higher in the <110> direction than <111> and <112> which could be the reason for the faster rate of expansion in the <110> direction that were documented in TEM research study and molecular dynamics/density functional theory (MD/DFT) [4]. Reprinted from AIP Advances 6, 125027 (2016) American Institute of Physics. Copyright by the author under the terms and conditions of the Creative Commons Attribution (CC BY 4.0).

## 5. Quantum Harmonic Oscillator

The computational model in this research work employs two conduction band or energy band quantum harmonic oscillator (QHO) where $E_0$ is the minimum conduction band and $E_1$ is the upper conduction band that is responsible for the optical amplification of the electromagnetic field [25]. The equations for the QHO are

$$E_{nls} = \left(n + \frac{1}{2}\right)\bar{h}\,\omega_{ij} + k_B T \ln\left[\frac{2}{n_c}\left(\frac{2\pi m_{eff} k_B T}{\bar{h}^2}\right)^{\frac{3}{2}}\right] \tag{21}$$

$$\omega_{ij} = \frac{e\vec{E}}{\bar{h}(3\pi^2 n_c)^{\frac{1}{3}}} \quad \text{for } n = 0 \tag{22}$$

$$\omega_{ij} = \frac{\bar{h}(3\pi^2 n_c)^{\frac{2}{3}}}{2m_{eff}} \quad \text{for } n = 1 \tag{23}$$

The indices in $E_{nls}$ are analogues to quantum numbers when describing atomic energy states. The index $n$ has the values 0 and 1, which label the minimum and upper conduction bands respectively. The index $l$ with values 1, 2 and 3, defines which angular momentum the energy state represents i.e., $l = 1$ is $\omega_{11}$, $l = 2$ is $\omega_{22}$ and $l = 3$ is $\omega_{33}$. The index $s$ is equal to $n_c$. The first term of Equation (21) describes the energy due to the angular momentum $\omega_{ij}$ of the electron. The second term is the thermal energy that the electron experiences while in the QHO where $k_b$ and $T$ are the Boltzmann constant and temperature, respectively. Each of the two energy bands are represented by a different angular momentum equation, as shown in Equations (22) and (23), and are composed of sub-energy levels or energy states that are a function of the negative charge differential $n_c$. It was found necessary to utilize two different angular momentums for $E_n$ in order to maintain the energy gap between the minimum and upper conduction bands and to prevent the energy states of $E_0$ and $E_1$ from over lapping. The QHO is nonlinear due to the angular momentum is not constant and is dependent of the negative charge differential $n_c$ [26,27]. The individual energy states in the conduction band $E_1$ increase exponentially in alternate patterns of the energy states that represents the three orthogonal directions of <110>, <111> and <112> (Figure 7a). The energy states are divided into two groups that depend on the magnitude of the drift velocity $v_d$. The allowable energy states have energies that have drift velocities below the speed of light $c$. All other energy states are above $c$ and therefore are disregarded.

The nonlinear QHO can be thought of as three interwoven quantum harmonic oscillators; one for each of the orthogonal directions. When all three directional nonlinear QHOs are analyzed separately, the individual energy states are shown to be not equally spaced (Figure 7b). Analyzing the $\mathrm{E}_1$ conduction band structure by separating the individual QHOs, one can see that most of the lowest energy states reside in the <110> direction. When a particular electron obtains enough energy due to photon absorption during optical amplification of the EM field, the electron transition from the minimum conduction band $E_0$ to an available energy state in $E_1$ (Figure 7c). It is the hypothesis of this research work that due to the strain that is created during the anisotropic volume expansion primarily in the <110> direction—due to optical amplification of the electromagnetic field—that the electrons in the sub-energy levels in $E_0$ will generally transition to the lowest and most stable energy states in the <110> direction in the $E_1$ conduction band since it has lowest sub-energy levels than the <111> and <112> energy state directions [28–30].

The minimum conduction band $E_0$ is also composed of a series of sub-energy levels or energy states in each of the three orthogonal directions the same as $E_1$. However, the individual energy gaps between the energy states in $E_0$ are smaller then $E_1$. Initially, when free electrons cross the interfacial reaction front from the crystallized silicon c-Si to the lithiated silicon $Li_xSi$, the electron flux has an average energy of 2 eV due to the 2 V energy source described in Figure 1. Most electrons that enter the lithium-silicon matrix only have enough energy to transition to the minimum conduction band $E_0$ since the lowest energy state in $E_1$ is 8.24 eV. Therefore, the probability that an electron will have enough energy to reach that energy state without optical amplification of the EM field is remote.

As a result, when electrons enter the lithiated silicon from c-Si, there is a high probability that they will transition to the minimum conduction band $E_0$ at which time photons are emitted that will aid in the spontaneous and stimulated emission process of lithium ions that was described earlier. Until optical amplification occurs, there is a low probability that any electrons will transition to the energy band $E_1$ and therefore the majority of electrons will exist in the $E_0$ energy band as they transition or 'bounce' from one energy state to another within the minimum energy band of $E_0$ (Figure 7d).

It is this movement of electrons within $E_0$ that creates the low applied electromagnetic field that was displayed in Figure 5b. The reason individual electrons transition from one energy state to another are the multiple electron–photon interactions within $E_0$. The lowest energy state in the minimum conduction band $E_0$ is 0.14 eV, which can be interpreted as the direct band gap energy between the maximum valence band and the minimum conduction band in lithiated silicon, which is in sharp contrast to crystalline silicon with an indirect band gap energy of 1.1 eV [31].

The nonlinear quantum harmonic oscillator within this computational model has similar features to the quantum confined Stark effect (QCSE) [32,33]. The nonlinear QHO can be thought of as a two-energy band quantum well, where upon an applied electric field would create a fine structure of separate energy states within both $E_0$ and $E_1$ conduction bands.

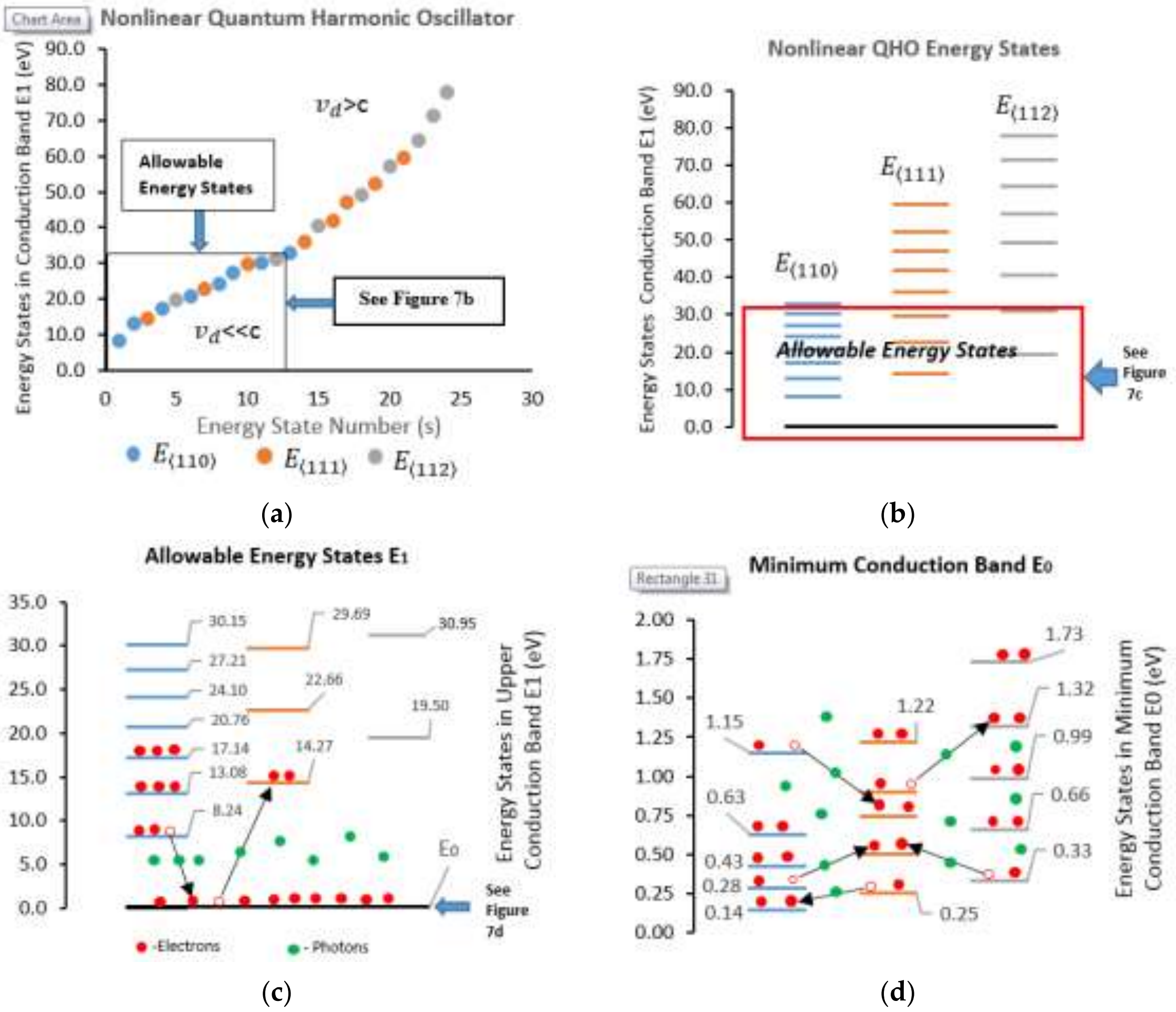

**Figure 7.** (**a**) The nonlinear quantum harmonic oscillator (QHO) has allowable energy states that are associated with electrons having drift velocities at non-relativistic speeds; (**b**) When analyzing the nonlinear QHO, it can be divided into three separate quantum harmonic oscillators, one for each of the three orthogonal directions of <110>, <111> and <112>; (**c**) The allowable energy states are displayed at a magnified view in order to examine the electron-photon interactions. When an electron absorbs a photon with sufficient energy by optical amplification the result is that the electron will transition from the minimum conduction band $E_0$ to $E_1$ and conversely when an electron emits a photon while in the upper conduction band the electron will transition from $E_1$ back to $E_0$; (**d**) A magnified view of the minimum conduction band $E_0$ showing the rapid transition of electrons as they move within $E_0$. The majority of electrons reside in the minimum conduction band and their multiple transitions in $E_0$ are the source of the low electromagnetic field prior to optical amplification as it was displayed earlier in Figure 5b.

## 6. Coherent State of Electrons

Since the free electrons are considered to exist in a discrete energy state within a quantum harmonic oscillator, these electrons can be thought of as being in a coherent state [34]. The Bloch function that was introduced in Equation (2) can be used to define the time-dependent coherent state of $E_1$

$$\mathrm{u_{cij}(r,t)} = \sum_{\mathrm{i=1}}^{3}\sum_{\mathrm{j=1}}^{3} \mathrm{U_{ij}} \exp\left[-\mathrm{i}\left(\mathrm{n}+\tfrac{1}{2}\right)\omega_{ij}\mathrm{t}\right]\mathrm{u_c(r)} \tag{24}$$

$$\mathrm{u_c(r)} = \mathrm{e^{ikr}} + \frac{1}{\mathrm{kr}}\mathrm{e^{i(\delta+kr)}}\sin\delta + \frac{3\mathrm{z}}{\mathrm{kr^2}}\mathrm{e^{i(\delta+kr)}}\sin\delta \tag{25}$$

where $i = j$, $n = 1$ and $\mathrm{U_{ij}}$ obeys the Poisson distribution

$$\mathrm{C_{ij}} = \left|\mathrm{U_{ij}}\right|^2 = \sum_{i=1}^{3}\sum_{j=1}^{3}\frac{\bar{n}_{eij}{}^{n_1} exp(-\bar{n}_{eij})}{n!} \tag{26}$$

with $\mathrm{C_{ij}}$ known as the electron probability. It is defined as the probability of an electron occupying the upper conduction band $E_1$ . The average number of electrons $\bar{n}_{eij}$ in the upper conduction band $E_1$ is defined as

$$\bar{n}_{eij} = Ia^3 = \frac{(\overleftrightarrow{E}_{ij}\times\overleftrightarrow{B}_{ij})a^3}{\mu_o\bar{\mathrm{h}}\omega_{ij}c}\mathrm{e}^{\gamma_{ij}\mathrm{r}} \tag{27}$$

which is the product of electromagnetic intensity $I(r)$ (Equations (14) and (15)) and the volume of the silicon cubic lattice [35]. The probability amplitude of $\mathrm{U_{ij}}$ is part of the quantum harmonic oscillator and therefore is in a dynamic state. This results in the time-dependent Bloch function $\mathrm{u_{cij}(r,t)}$ of being a function of the average negative charge differential $\bar{n}_c$.

The location of the probability amplitudes $\mathrm{U_{ij}}$ for each of the coherent states for the three orthogonal directions of <110>, <111> and <112> corresponds to the maximum EM field energy as displayed in Figure 5a. When $\bar{n}_c = 6$, direction <110> has a maximum EM energy of approximately 30 eV and a $C_{\langle 110\rangle}$ of 0.3553. Similarly, for the <111> ($\bar{n}_c = 13$) and <112> ($\bar{n}_c = 21$) directions, we have the corresponding $C_{\langle 111\rangle} = 0.2546$ and $C_{\langle 112\rangle} = 0.3062$ respectively, with an EM energy of approximately 30 eV (Figure 8).

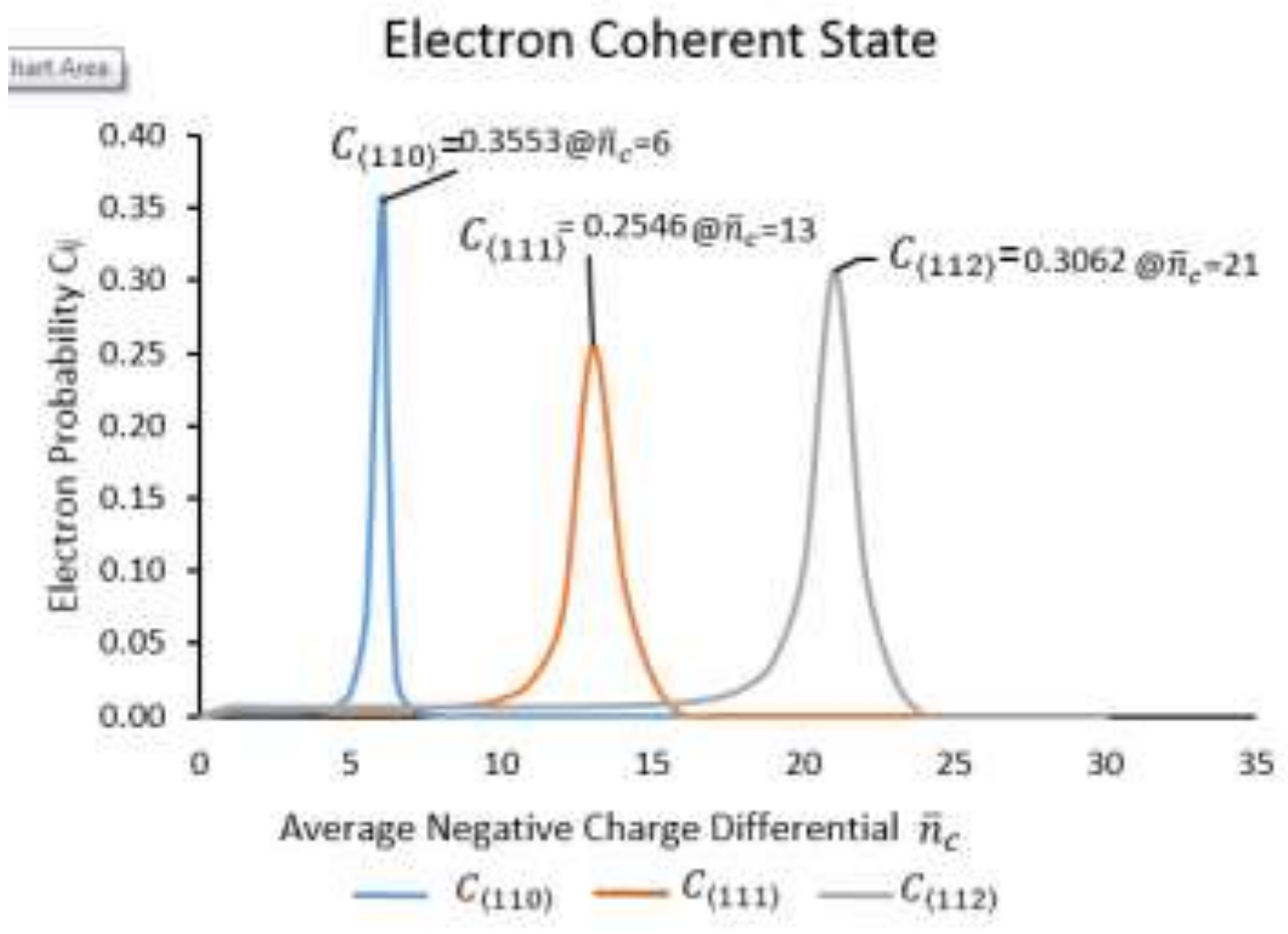

**Figure 8.** The highest probability that an electron might transition from the minimum conduction band $E_0$ to the upper conduction band $E_1$ corresponds to a specific average negative charge differential $\bar{n}_c$, where the electromagnetic field energies are at their maximum for each of the three orthogonal directions.

## 7. Photon Density Wave

The applied electromagnetic field, as defined with the aid of Maxwell stress and optical amplification, is mainly a quantum mechanical theory. The photon density wave theory will be utilized to explain the absorption rate of the individual photons that comprise the EM field. Originally this theory is not based on quantum mechanics, but on a diffusion theory based on a partial differential equation, that is different from the Schrödinger equation. However, this theory will be transformed from a diffusion theory, where the photons are thought of as material-like particles diffusing through a medium, to that of a quantum field theory, where the photons are the traditional particles of energy. The insertion of Fermi's golden rule is the quantum mechanical equation that is used to define the photon absorption rate of silicon atoms [36]. The group of equations for the photon density wave theory is

$$\nabla^2 I_{ij}(r,t) - \frac{c\mu_{aij}}{D_{\gamma ij}} I_{ij}(r,t) - \frac{1}{D_{\gamma ij}}\frac{\partial I_{ij}(r,t)}{\partial t} = -\frac{1}{D_{\gamma ij}} S_{ij}(r,t) \quad (28)$$

$$I_{ij}(r,t) = \frac{\bar{n}}{a^3} = \frac{\vec{E}_{ij} \times \vec{B}_{ij}}{\mu_o \hbar \omega_{ij} c} e^{(\gamma_{ij} r - \omega_{\gamma ij} t)} \quad (29)$$

$$S_{ij}(r,t) = c\mu_{aij} I_{ij} \quad (30)$$

$$D_{\gamma ij} = \frac{c}{3\left(\mu'_{sij} + \mu_{aij}\right)} \quad (31)$$

$$\mu'_{sij} = \frac{1}{l_{eij}}(1 - \langle \cos\delta \rangle) \quad (32)$$

$$\mu_{aij} = \frac{W_{\gamma ij}}{\bar{n}_{eij}} \quad (33)$$

$$W_{\gamma ij} = \frac{2\pi}{\hbar} \left|\langle \mathrm{u}_c | H_p | \Psi^g_{Si} \rangle\right|^2 \frac{1}{4\pi^2}\left(\frac{2m_{eff}}{\hbar^2}\right)^{\frac{3}{2}} \left(\mathbb{E}_{ij} - \mathbb{E}_{coh}\right)^{\frac{1}{2}} \quad (34)$$

where $\mathbb{E}_{ij} \geq \mathbb{E}_{coh}$ [37].

The general solution to Equation (28) is the time-dependent electromagnetic intensity $I(r,t)$ as stated in Equation (29). $S_{ij}(r,t)$ is defined as the source of the electron flux, $D_{\gamma ij}$ is the photon diffusion coefficient and $\mu'_s$ is the reduce scattering coefficient, which is a function of the phase shift $\delta$ and mean free path $l_e$ of the number of electrons in the two conduction bands. The absorption coefficient $\mu_{aij}$ is defined by the photon absorption rate of silicon atoms $W_{\gamma ij}$ , $\bar{n}_{eij}$ is the mean number of electrons within the EM field, $\Psi^g_{Si}$ is the silicon atom ground state wave function and $\mathbb{E}_{coh}$ is the cohesive energy between two silicon atoms [38]. The applied electromagnetic field has photons of varies energies. Once again, it is the hypothesis of this research that the photons must have energies greater than the cohesion energy in order to break any diatomic silicon bond for volume expansion to begin. The boundary conditions for the diffusion partial differential equation of Equation (28) are defined by the Fresnel coefficients $F_{\gamma ij}$

$$\frac{n^2_{1ij}}{n^2_{0ij}} I_{0ij}(r,z)|_{z=0} - F_{\gamma ij} D_{0ij} \frac{\partial I_{0ij}(r,z)}{\partial z}|_{z=0} = I_{1ij}(r,z)|_{z=0} \quad (35)$$

$$n_{\gamma ij} = \sqrt{\varepsilon_{ij}\mu_{ij}}, \qquad \mathrm{i = j = 1\ or\ 2\ or\ 3}, \qquad \gamma = 0 \text{ or } 1, \quad (36)$$

where the refractive index $n_{\gamma ij}$ is the same as $n_{ij}$ as defined in Equation (20). The index $\gamma$ indicates the difference in the adjacent medium that the photon experiences as it travels through the lithiated silicon matrix. The value of $\gamma$ = 0 is the initial part of the medium, that the photon travels through, where $\gamma$ = 1 is the adjacent part of the medium in which the photon experiences. The difference of the two media, since both are heterogeneous lithium-silicon materials, is the difference in the negative charge differential $n_c$ in both media. When boundary conditions are satisfied, the Fresnel coefficients $F_{\gamma ij}$ are used to calculate the photon transmission $T_{ij}$ and the photon reflection $R_{ij}$ coefficients [39].

$$T_{ij} = \frac{2{n_{1ij}}^2 D_{0ij} k_0}{n_{0ij}^2 D_{0ij} k_0 \left[1 - iF_{\gamma ij} D_{1ij} k_1\right] + k_1 D_{1ij} n_{1ij}^2} \quad (37)$$

$$R_{ij} = \frac{k_0 D_{0ij} n_{0ij}^2 [1 - iF_{\gamma ij} D_{1ij} k_1] - k_1 D_{1ij} n_{1ij}^2}{k_0 D_{0ij} n_{0ij}^2 [1 - iF_{\gamma ij} D_{1ij} k_1] + k_1 D_{1ij} n_{1ij}^2}, \quad (38)$$

where $k_0$ is the photon incident wave vector and $k_1$ is the photon scattering wave vector. These two coefficients are interpreted in terms of quantum mechanics and is analogues to transmission and reflection coefficients of the probability current density. When the photon experiences both media with the same $n_c$ value, the refractive indices are $n_{0ij} = n_{1ij}$; therefore $T_{ij} = 1$ and there is no silicon atom photon absorption in this part of the media. Please note, that the photon density wave theory is originally a classical or continuum wave theory and as such the incident wave on a surface reflects according to a prescribed angle. However, since in this quantized version of this theory, the waves are actually photons, the reflection coefficient $R_{ij}$ acts like an absorption coefficient with values between 0 and 1 (Figure 9).

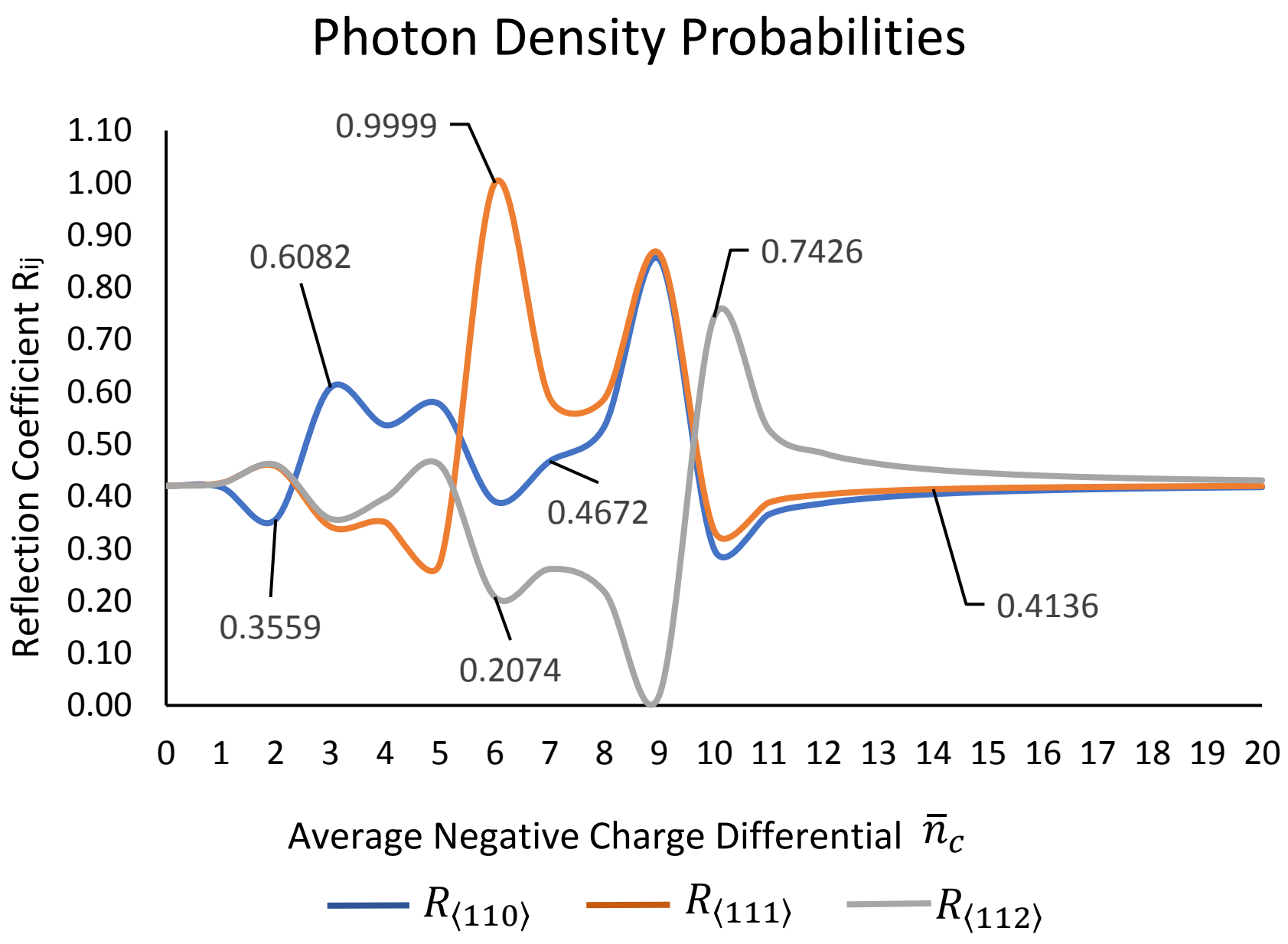

**Figure 9.** The reflection coefficient $R_{ij}$ is the probability that after a photon is emitted during an electron transition from the $E_1$ conduction band to the minimum conduction band $E_0$ that the photon will be absorbed be a silicon atom.

## 8. Path Integral Method

The transition probability $P_{ij}$ is the product of three probability variables that have been defined earlier in this study, namely the negative charge differential probability $N_{pq}$, the electron probability $C_{ij}$ and the photon reflection coefficient $R_{ij}$

$$P_{ij} = N_{pq} C_{ij} R_{ij} \quad (39)$$

The transition probability $P_{ij}$ is the probabilistic calculation of a series of events that starts with the probability $N_{pq}$ (that a specific average negative charge differential $\bar{n}_c$ will exist at any given moment), which follows the probability $C_{ij}$ (that an electron might absorb a photon and transition from the minimum conduction band $E_0$ to the energy band $E_1$ at a specific energy state) and finally the probability $R_{ij}$ (that an electron might transition from the $E_1$ energy band back to the minimum conduction band $E_0$ and in so doing emit a photon that will be absorbed by a silicon atom). Figure 10 displays the transition probabilities of $P_{11}$, $P_{22}$ and $P_{33}$ for each of the three orthogonal directions <110>, <111> and <112>, respectively. As it can be examined, each transition probability has a Poisson distribution that is similar to one another with amplitudes ranging approximately between 0.020 and 0.025. The inference of the three directional $P_{ij}$ distributions is that the expansion rate should be closer to being isotropic in nature instead of the anisotropic volume expansion that is actually witnessed in TEM images [40].

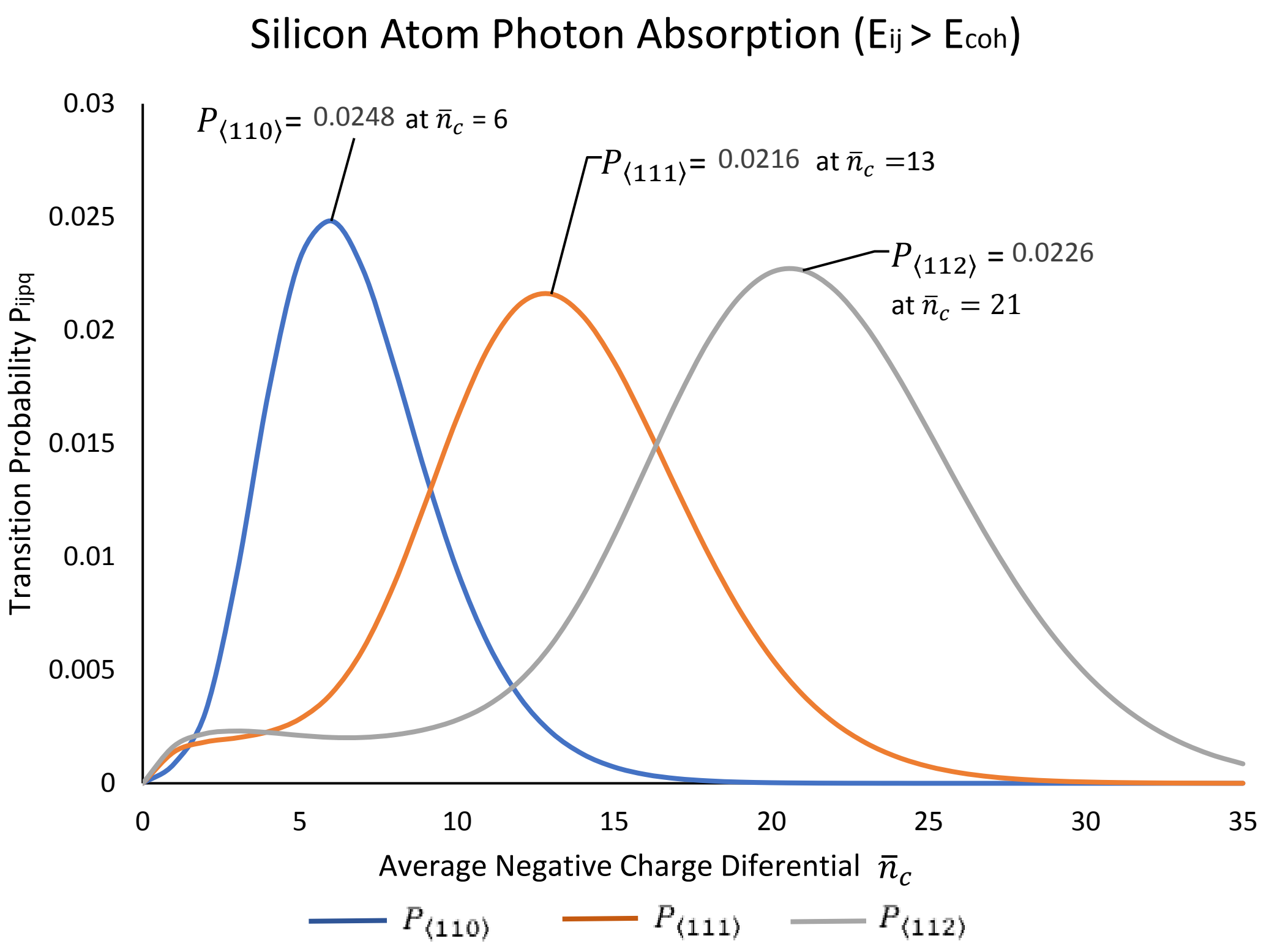

**Figure 10.** The transition probability $P_{ij}$ for each of the orthogonal directions suggest that the expansion rate and volume increase is resembling to be isotropic rather than anisotropic as reported in several research studies.

In order to derive the anisotropic geometry of the silicon nanowire during lithium ion insertion the transition probability is interpreted as a propagator $P_{ij}$, which is an important element of the path integral method in quantum field theory [41]

$$P_{ij} = a\sqrt{\frac{m_{eff}\omega_{ij}}{2\pi i\bar{h}}}\,e^{i\frac{S_{ij}}{\bar{h}}}, \tag{40}$$

where $S_{ij}$ is the principle of least action defined for this paper as:

$$S_{ij} = \int_{\tau_0}^{\tau_1} \Delta\mathbb{E}_{ij} dt = \int_{\tau_0}^{\tau_1} \vec{p}_{ij}\cdot\Delta\dot{\vec{r}}_{ij}\,dt = -i\bar{h}\vec{\nabla}_{ij}\int_{\tau_0}^{\tau_1}\Delta\dot{\vec{r}}_{ij}\,dt = -i\bar{h}\ln(\tau_{ij})\Delta\vec{r}_{ij}\vec{\nabla}_{ij}. \tag{41}$$

The change in energy $\Delta\mathbb{E}_{ij}$ is the change in the Maxwell stress tensor, $\vec{p}_{ij} = -i\bar{h}\vec{\nabla}_{ij}$ is the momentum operator of the electron, $\Delta\vec{r}_{ij}$ is the change in distance of the position vector $\vec{r}$ that will be defined below and $\tau_{ij}$ is known as the time-normalization defined as:

$$\tau_{ij} = \frac{a m_{eff}\omega_{ij}}{e\vec{\mathrm{E}}_{ij}t}. \tag{42}$$

The time-normalization $\tau_{ij}$ is dimensionless and due to the drift velocity $v_d$ of the electron flux [42]. The definition of $\tau_{ij}$ is the amount of time it takes for the electrons in $n_c$ to travel the distance of the silicon lattice constant $a$. In general, the propagator $P_{ijpq}$ can best be described mathematically as a function that transforms an initial state of a wave function $\Psi_i$ to a final state $\Psi_f$

$$\Psi_f(\vec{r}_{1ij}, t_1) = \int_{-\infty}^{\infty} \Psi_i(\vec{r}_{0ij}, t_0) P_{ij}(\Delta\vec{r}_{ij}, \Delta t)\, dr, \tag{43}$$

where $\vec{r}_{0ij}$ is the initial state vector of $\Psi_i(\vec{r}_{0ij}, t_0)$ and $\vec{r}_{1ij}$ is the final state vector of $\Psi_f(\vec{r}_{1ij}, t_1)$. The propagator $P_{ij}$ will transform the lithium ion excited state wave function $\Psi_{Li}^{e}$ defined earlier as

$$\Psi_{Li}^{e} = \frac{\langle \Psi_{Li}^{m} | H_p | \Psi_{Li}^{g} \rangle}{(E_o - E_m)} \Psi_{Li}^{m} \tag{44}$$

Before we continue with the path integral method, it is necessary to further refine the definition of $n_c$. Prior to the electrons crossing the interfacial reaction front and encountering lithium ions, the positive Li ions are the only charge particles within the lithiated silicon lattice structure (Figure 11a). These positive-charge particles are under a Coulombic repulsive potential that intends to push them farther apart from one another. When the electron flux encounters these Li ions, the negative electrons interact with the positive Li ions, in which an attractive potential manifests that creates the quasi-particles called polarons [9]. Each individual polaron will be assigned the variable $n_c$ , for which this value changes with respect to both space and time. The polaron are define by a wave function $\Psi_p^e$ and a transition state vector $\vec{r}_{ij}$ which is define as the difference between initial and final state vectors $\vec{r}_{ij} = \vec{r}_{1ij} - \vec{r}_{0ij}$ (Figure 11b).

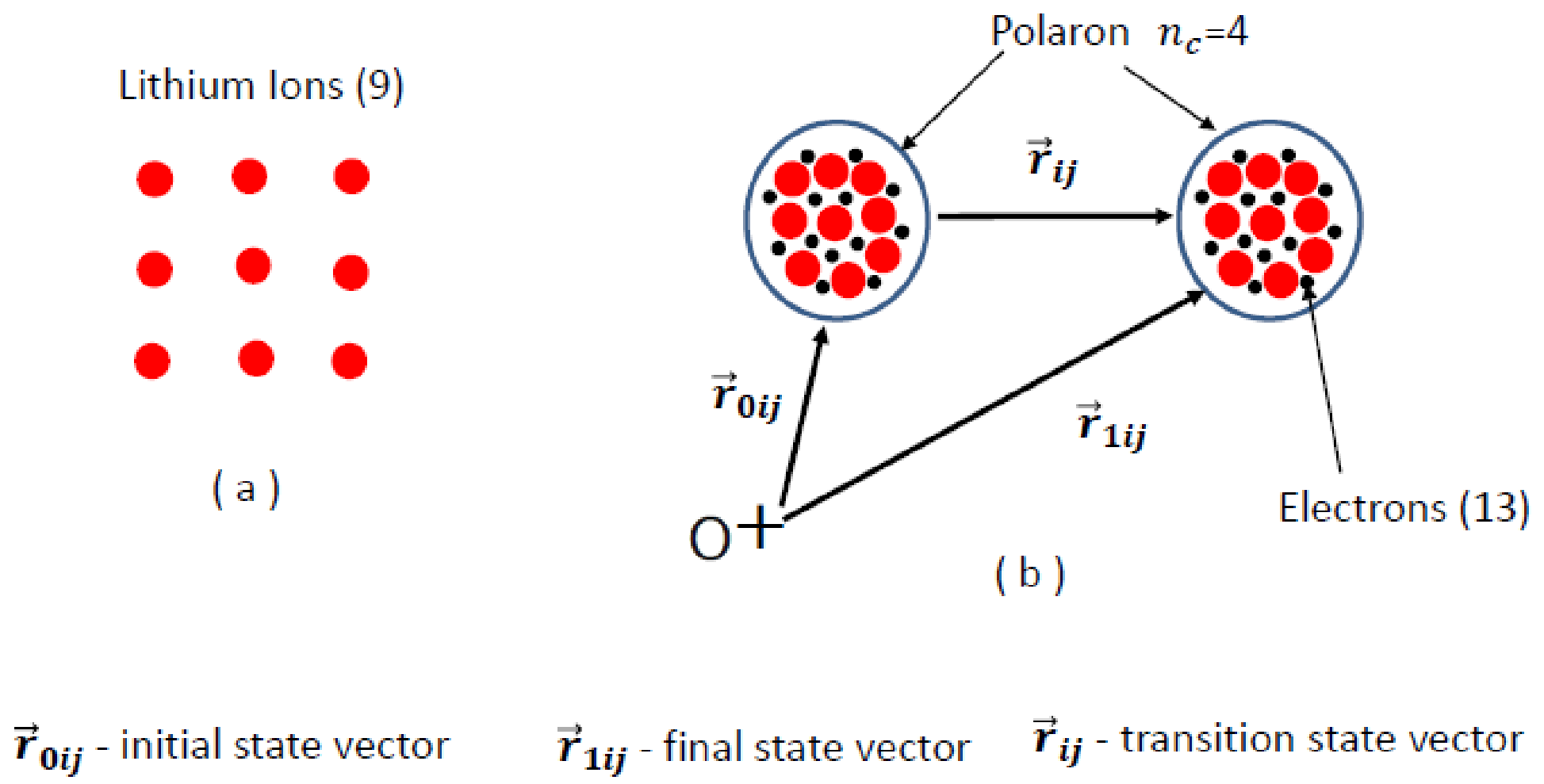

**Figure 11.** (**a**) Lithium ions before they encounter the electron flux; (**b**) The lithium ion and electrons form a quasi-particle known as a polaron, which in this computational model has a surplus of electrons compared to the positive ions and therefore a negative charge differential of $n_c$. In the example there are 9 Li ions and 13 electrons, which create a polaron of $n_c$ = 4.

In this research study an assumption is established that $\Psi_{Li}^{e}$ is similar to the polaron wave function $\Psi_p^e$, therefore $\Psi_p^e \approx \Psi_{Li}^{e}$ . Since both wave functions consist of the same constitutive particles that define the same mathematical parameters within their respective wave functions, i.e., the quasi-particle polarons are composed of the excited state Li ions and free electrons within the cubic silicon lattice. The reason the Li ion is in the excited state is indirectly caused by energetic free electrons emitting photons that cause excitation of the Li ions. The initial state of $\Psi_p^e$ is at the initial state vector $\vec{r}_{0ij}$ and the final state of $\Psi_p^e$ is at the final state vector $\vec{r}_{1ij}$. With further algebraic manipulation of Equations (40) and (41) and subsequently introducing $\Psi_p^e$ into Equation (40), the path integral equation becomes

$$\Delta\vec{r}_{ij}\vec{\nabla}_{ij}\Psi_p^e = \tau_{ij}\left(\frac{P_{ij}}{a}\sqrt{\frac{2\pi i\hbar}{m_{eff}\omega_{ij}}}\right)\Psi_p^e. \tag{45}$$

This path integral equation is a solution for $\Delta\vec{r}_{ij}$ which is defined as the decrease in the length of the transition state vector $\vec{r}_{ij}$ (Figure 12). The reason for the decrease in the transition state vector is greater in the <110> direction than the <111> and <112> direction is due to a larger electromagnetic field energy cause by optical amplification as described earlier in Figure 6b.

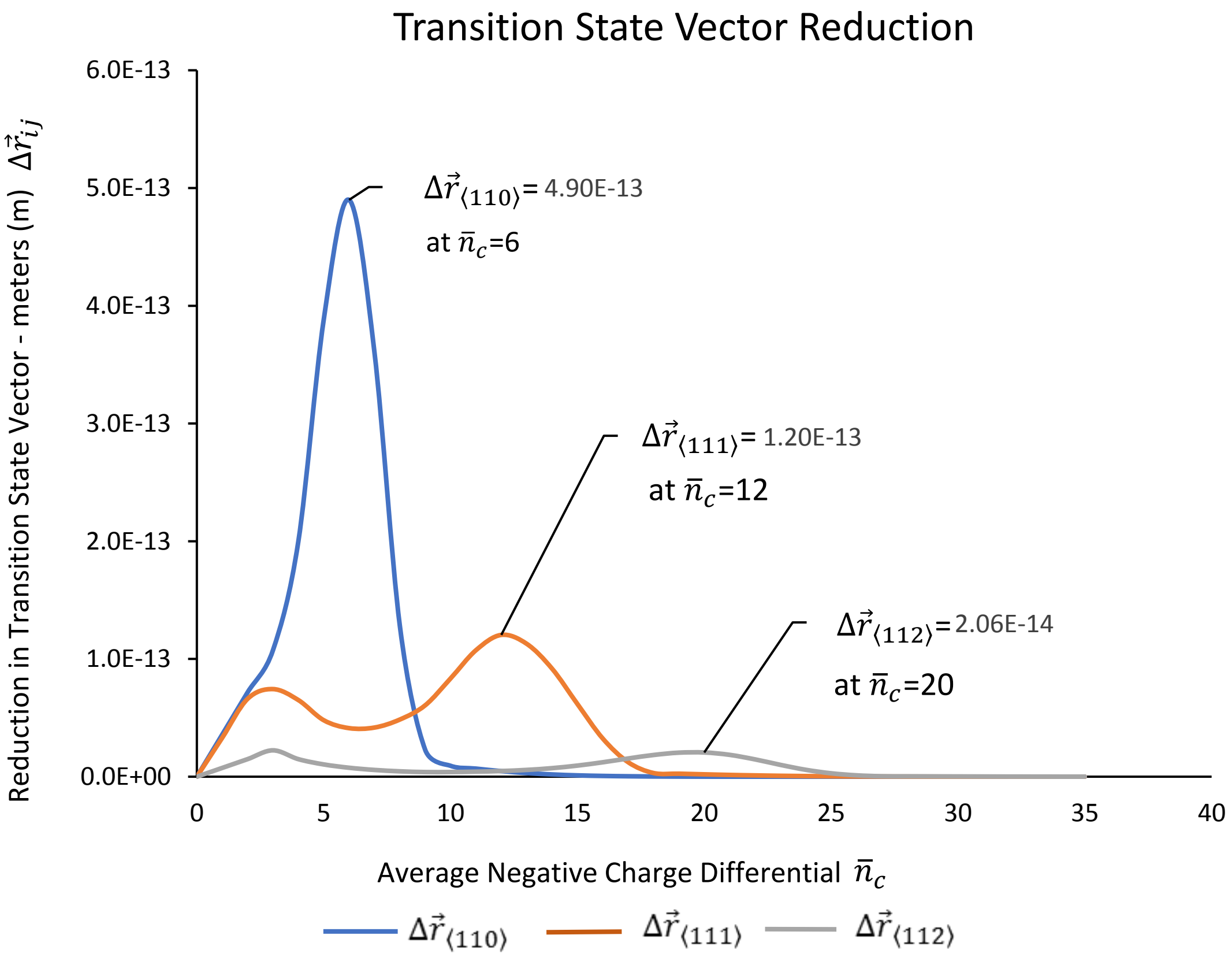

**Figure 12.** The decrease in the transition state vector $\Delta\vec{r}_{ij}$ is result of the amplification of the electromagnetic field, which is greater in the <110> orthogonal direction than <111> and <112>.

As a polaron transitions from an initial state arrive at a final state (Figure 11), the transition probability $P_{ij}$ at the end of the transition state vector $\vec{r}_{ij}$ is the parameter that determines whether an electron from the polaron will emit a photon $\mathbb{E}_{ij}$ that will be absorb by a silicon atom with energy greater than the cohesive energy $\mathbb{E}_{coh}$ of two silicon atoms. In general, the shorter the transition state vectors $\vec{r}_{ij}$ in one of the orthogonal directions, the greater the number of transition probability $P_{ij}$ events per unit length in that direction.

The transition probability $P_{ij}$ for each of the three orthogonal directions is very low in magnitude as shown previously in Figure 10. Therefore, at each final state vector there is a low probability that there will be a photon absorption event of $\mathbb{E}_{ij} \geq \mathbb{E}_{coh}$. When there is an equal number of Li ions and electrons ($n_c = 0$) in a cubic lattice structure, the length of the transition state vector $\vec{r}_{ij}$ is the same in all three orthogonal directions. With the increase in $\bar{n}_c$ and thus the increase in the applied electromagnetic field, the $\vec{r}_{ij}$ vector decreases in length. The differences between the three transition state vectors $\vec{r}_{11}$, $\vec{r}_{22}$, $\vec{r}_{33}$ in the <110>, <111> and <112> directions respectively become very significant when optical amplification develops (Figure 13). The largest reduction in vector $\vec{r}_{11}$ is at $\bar{n}_c = 6$ when the applied electromagnetic field is at its maximum in the <110> direction, with only minimal and negligible decrease changes in vectors $\vec{r}_{22}$ and $\vec{r}_{33}$, respectively.

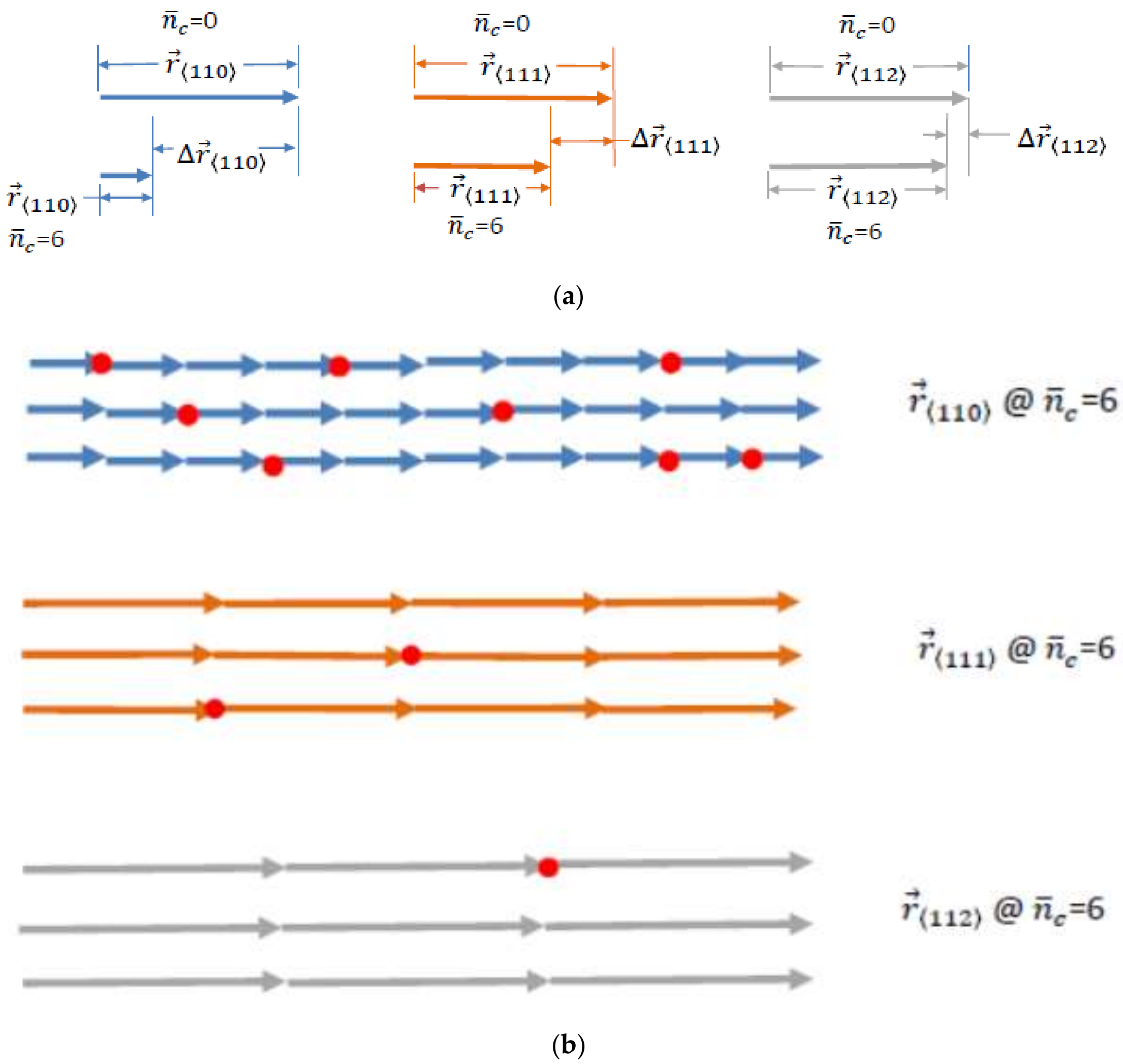

**Figure 13.** (**a**) Comparing the three transition state vectors before and after optical amplification; (**b**) A series of transition state vectors are linked together when $r_{ij}$ is at $\bar{n}_c = 6$. The red dots mark the events when photon absorption occurs when the applied electromagnetic energy is greater than the silicon atom cohesion energy.

At the end of each transition state vector $\vec{r}_{ij}$ at the final state there is a probability for an occurrence or an event. In this work, there are four probable events that could occur following the emittance of a photon due to the transition of an electron from $E_1$ conduction band to the minimum conduction $E_0$. The first probability is that the photon could interact with an electron that could possibly give the electron sufficient energy to transition to a higher energy state. Another probability is that the photon could be absorbed by a Li ion, which add in the optical amplification process and augment the electromagnetic field. The third probability is that the photon goes through a transmission process travel threw silicon lattice model without being absorbed. The fourth and final possible occurrence is the only event that this research study is interested in—the probability of the photon being absorbed by a silicon atom.

## 9. Anisotropic Expansion Rate

The expansion rates are compared from the $\Delta\vec{r}_{ij}$ calculations that were determined from the path integral method in Figure 12. The ratio of $\Delta\vec{r}_{11}/\Delta\vec{r}_{22}$ (or $\Delta\vec{r}_{<110>}/\Delta\vec{r}_{<111>}$) fastest expansion rates was calculated to be 11.88 at $\bar{n}_c = 6$ which is greater than what was determine from the density functional theory study, which reported an expansion rate seven times faster in the <110> direction than in the <111> direction [4].

However, the expansion rate that is reported in this study correlates with a TEM research study, where the finding implicitly determined the anisotropic expansion rate was between 10.5 and 13.3 greater expansion in the <110> direction than the <111> direction [43]. The other two orthogonal directions fastest anisotropic volume expansion rates were calculated to be $\Delta\vec{r}_{<110>}/\Delta\vec{r}_{<112>} = 64.57$ and $\Delta\vec{r}_{<111>}/\Delta\vec{r}_{<112>} = 25.20$ at $\bar{n}_c = 6$ and $\bar{n}_c = 11$ respectively (Figure 14).

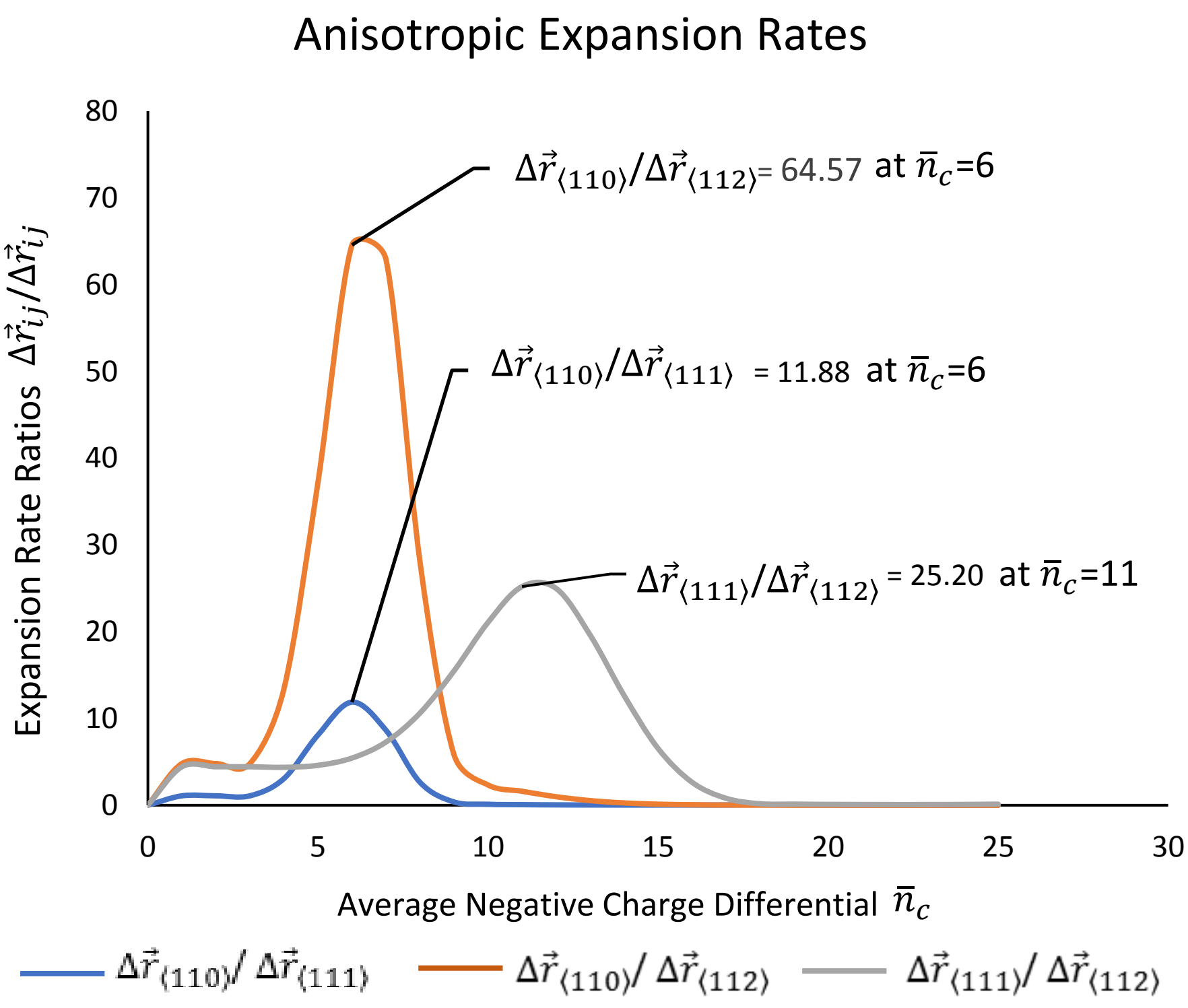

**Figure 14.** The orthogonal expansion rate ratio in the $\Delta\vec{r}_{\langle 110\rangle}/\Delta\vec{r}_{\langle 111\rangle}$ directions corresponds to that which was reported by previous research studies both experimentally with the transmission electron microscopy and with density functional theory /molecular dynamics simulations.

## 10. Volume Increase and Geometric Cross-Sectional Area Configuration

The volume expansion and geometric configuration of the silicon nanowire at full lithiation can best be demonstrated by deriving a set of equations from the Cassini oval geometry [40]. In this study, the computational model simulated the nanowire volume increase of 300% upon the conclusion of lithium ion insertion. There is a volume change $\Delta V_{ij}$ in each of the three orthogonal directions of <110>, <111> and <112> , with each volume component a function of $\Delta\vec{r}_{ij}$

$$\begin{gathered}\Delta V_{11} = V_{max}\frac{\Delta\vec{r}_{11}^{2}}{\Delta\vec{r}_{max}^{2}}\\ \Delta V_{22} = V_{max}\frac{\Delta\vec{r}_{22}^{2}}{\Delta\vec{r}_{max}^{2}}\\ \Delta V_{33} = \frac{\Delta\vec{r}_{33}}{\Delta\vec{r}_{max}}\sqrt{\left(\frac{2}{\pi}V_{max}\right)}\end{gathered} \tag{46}$$

with $V_{max}$ being the total maximum volume increase and is define as $V_{max} = \Delta V_{11} + \Delta V_{22} + \Delta V_{33}$ (Figure 15a). The maximum decrease in the transition state vector $\Delta\vec{r}_{max}$ is defined as $\Delta\vec{r}_{11}$ at $\bar{n}_c$ = 6 in the <110> direction as displayed in Figure 12.

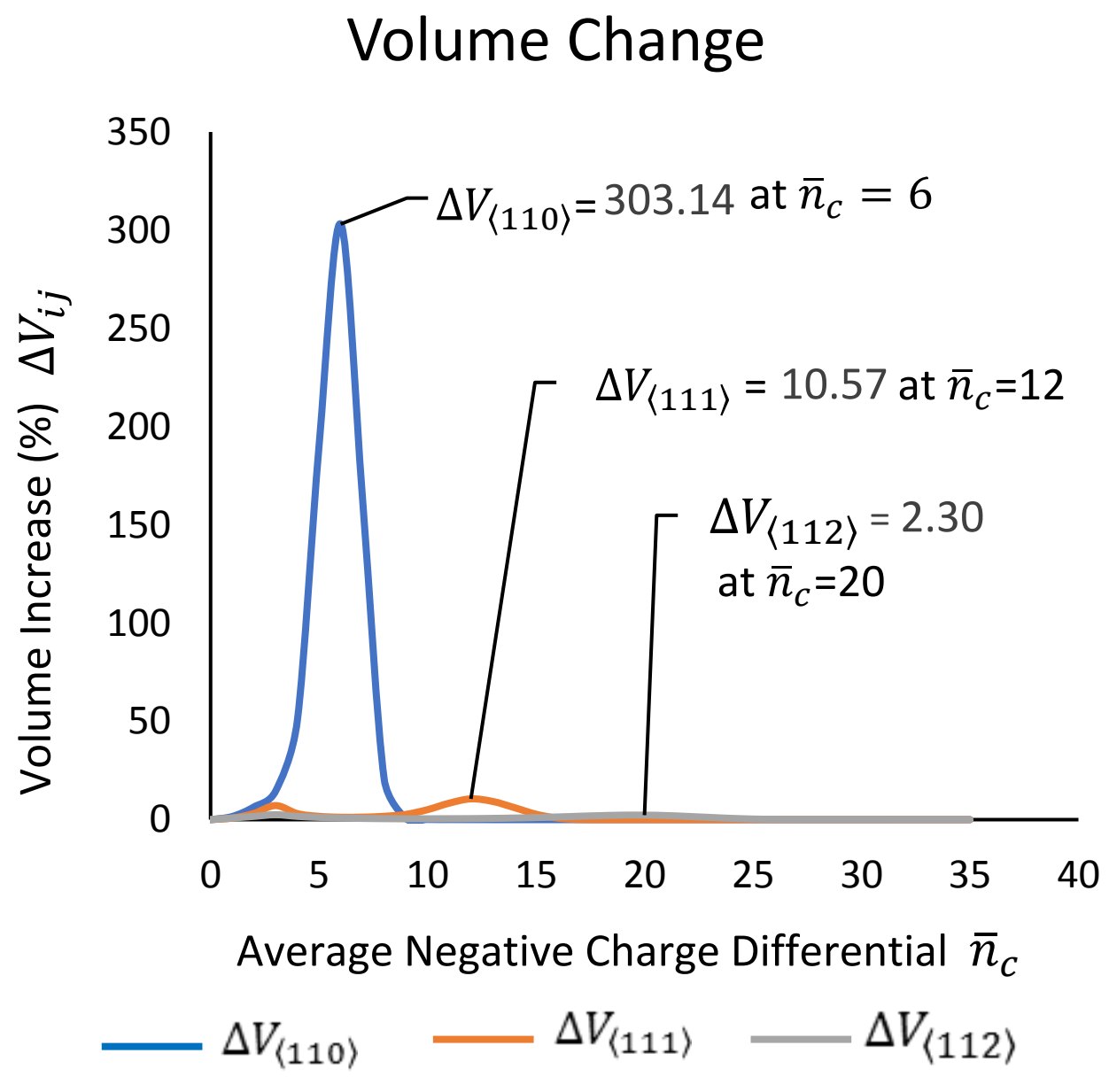

(**a**)

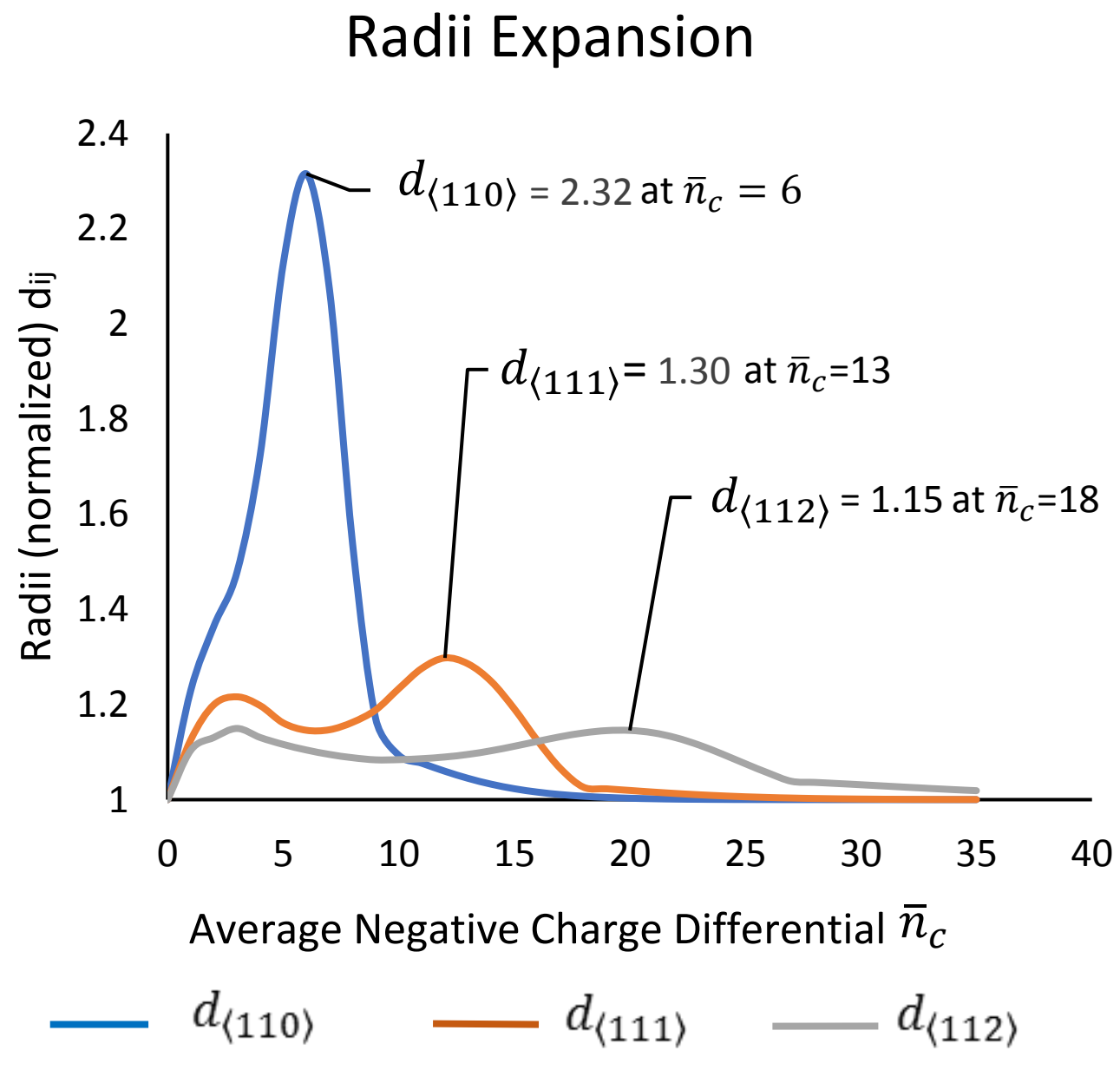

(**b**)

**Figure 15.** (**a**) The computational model simulated a maximum volume increase of over 300% during full lithiation. The majority of the volume increase was in the <110> direction with only a small volumetric increase in the <111> and <112> directions; (**b**) The largest radius increase was in the <110> direction due to the overwhelming volume expansion in this direction compared to the <111> and <112> orthogonal directions.

The Cassini oval mathematics was applied, because the cross-sectional area of the fully lithiated silicon nanowire resembles the analytical geometry of those set of mathematical curves. The radii $d_{ij}$ of the cross-sectional area are equations that are also derived from the Cassini oval geometry and are defined as

$$d_{11} = \beta_{11}\left(V_{max}\frac{\Delta\vec{r}_{11}^{2}}{\Delta\vec{r}_{max}^{2}}\right)^{\frac{1}{3}}$$

$$d_{22} = \beta_{22}\left(V_{max}\frac{\Delta\vec{r}_{22}^{2}}{\Delta\vec{r}_{max}^{2}}\right)^{\frac{1}{3}} \quad (47)$$

$$d_{33} = \beta_{33}\left(\frac{\Delta\vec{r}_{33}}{\Delta\vec{r}_{max}}\sqrt{\left(\frac{2}{\pi}V_{max}\right)}\right)^{\frac{1}{3}},$$

where the constant $\beta_{ij}$ is based on the orientation between the three orthogonal directions and the reference coordinate system of the computation model (Figure 15b). The radius value of $d_{ij} = 1$ in all three orthogonal directions defines the initial volume of the silicon nanowire prior to anisotropic expansion. The lithiated silicon nanowire is at its largest volume expansion when the average negative charge differential is $\bar{n}_c$ = 6. This corresponds to the time when the radius $d_{\langle 110\rangle}$ is at its greatest value $d_{11} =$ 2.31 and the radius $d_{\langle 111\rangle}$ is at $d_{22}$ = 1.10, which together these two radii form the Cassini oval cross-section geometry (Figure 16). This specific shape has been witnessed in TEM before. However, this work predicts that at $\bar{n}_c$ = 13 the radius in the <111> direction is slightly larger ($d_{22}$ = 1.29) than in the <110> direction ($d_{11}$ = 1.10). At even higher $\bar{n}_c$ values ($\bar{n}_c$ >18), there is a negligible increase in volume in the cross sectional area of the silicon nanowire with a very small increase in volume in longitudinal direction of <112> with an increase in length of $L_{33} = 1.15$ at $\bar{n}_c = 18$.

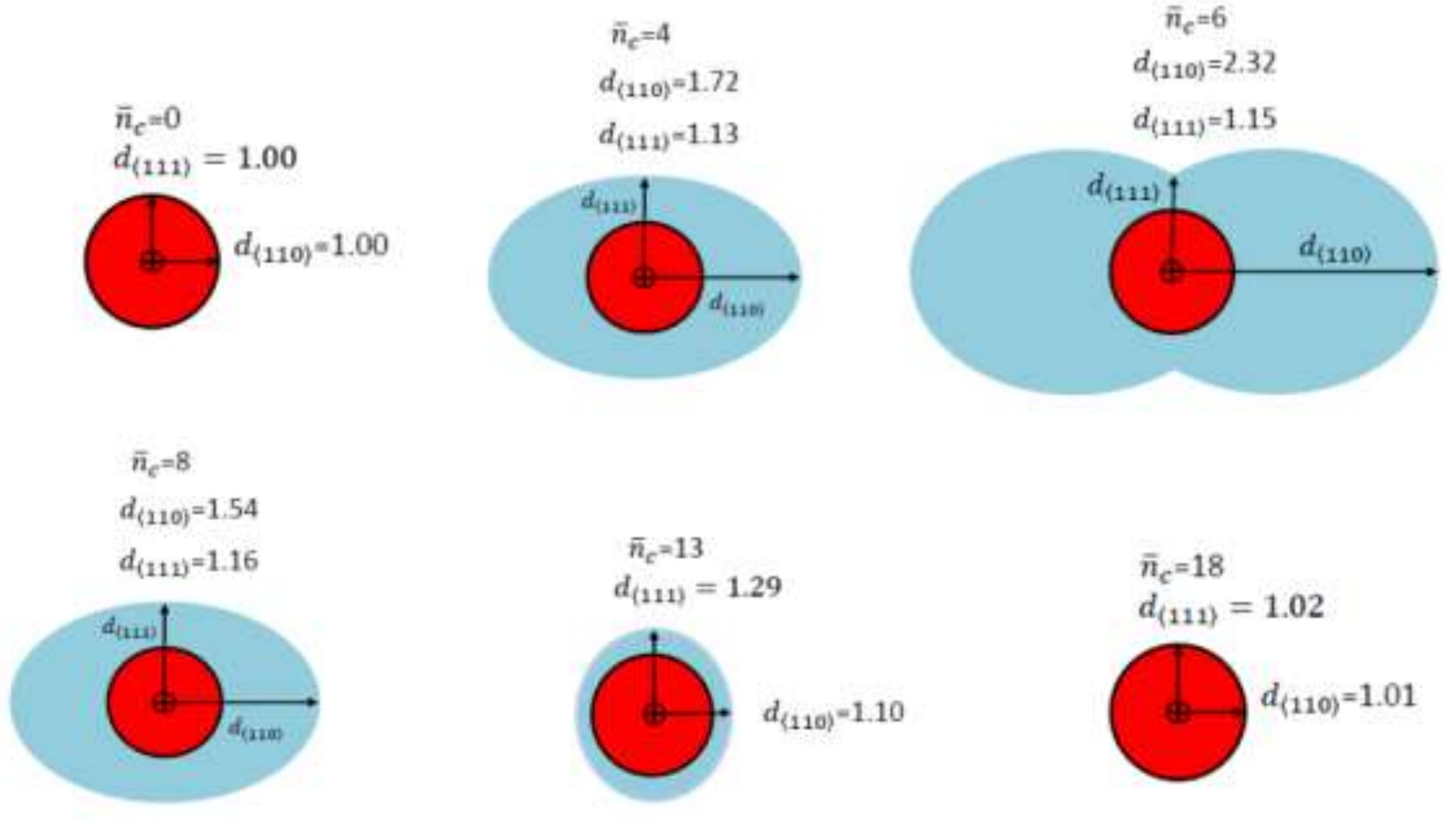

**Figure 16.** The cross-sectional configuration at varies $\bar{n}_c$ values. As the average negative charge differential increases, the anisotropic Cassini oval area reaches its peak at $\bar{n}_c$ = 6. At this point with increasing $\bar{n}_c$ the cross-sectional areas are beginning to resemble areas from lower $\bar{n}_c$. It is important to note that the lithiated silicon nanowire is not elastic however plastic in nature. Therefore, once the nanowire has expanded to a certain volume, it does not contract to a previous smaller volume [2,3,44].

## 11. Summary

In this work, we have demonstrated the need for a multi-physics approach based on first principle theories in order to create a computational model to study the quantum effects of a silicon nanowire during lithium ion insertion. The working hypothesis that has been presented in this paper begins with an electron flux, lithium ions and silicon atoms that serve as the background within our model. The net negative charge density is defined by an independent variable called the negative charge differential $n_c$ and is defined mathematically by a Poisson distribution. This $n_c$ variable ultimately defines the energy within our model as the moving charges induce an applied EM field. The energy of this EM field is defined by the Maxwell stress tensor.

With the application of energy from the applied electromagnetic field, several processes begin. First, the applied EM field begins the process of spontaneous and stimulated emission that will lead to the optical amplification of the electromagnetic field. This will ultimately lead to a coherent state for the electrons once amplification occurs. With the increase of the electromagnetic field, energetic photons within the field interact with constitutive particles that are present within the matrix of our model. These photons will be absorbed by electrons, lithium ions and silicon atoms. The optical amplification of photons will be absorbed by electrons in the minimum conduction band and transition to the upper conduction band. In an iterative process, amplified photons that are absorbed by lithium ions—which were induced by optical amplification initially—will continue to be the driving energy throughout the lithiated silicon nanowire. In addition, amplified photons with energy greater than the cohesion energy of silicon atoms will also be absorbed by these particles that will possibly break the cohesion bond between two silicon atoms that will initiate the start of silicon nanowire expansion.

It was also demonstrated in this research that true to the nature of quantum physics, physical processes are governed by probabilistic events. A series of calculated events culminated into the transition probability $P_{ij}$ , that stochastically determines the silicon atom photon absorption rate. However, that alone would not produce the anisotropic expansion that is witnessed in TEM images. The optical amplification of the EM field was discovered to be greater in the <110> orthogonal direction than in the <111> and <112> direction due to the refractive indices $n_{ij}$ or electric susceptibilities $\varepsilon_{ij}$ of the lithium-silicon material. The $n_{ij}$ or $\varepsilon_{ij}$ were present in the Maxwell stress, the optical amplification, the nonlinear quantum harmonic oscillator, the electron coherent state, the photon density waves and the path integral method equations. The refractive indices were found to be inversely proportional to the electric field and thus as the electric susceptibilities decrease in magnitude the electric field increases in strength especially in the <110> direction. The path integral method was the mathematical theory to show that a dramatic increase in the Maxwell stress or energy density in the <110> orthogonal direction will result in a greater expansion in this direction than in the <111> and <112> orthogonal directions.